\pdfoutput=1
\documentclass[PAPER, atlasdraft=false, coverpage=false, cernpreprint=true, UKenglish, texmf, orcidlogo]{atlasdoc}
\usepackage{atlaspackage}
\usepackage{atlasbiblatex}

\usepackage{atlasphysics}
\graphicspath{{logos/}{figures/}}

\usepackage{ANA-HIGP-2024-49-PAPER-defs}
\usepackage{multirow}

\AtlasTitle{Cross-section measurements of boosted Higgs boson production in final states with pairs of hadronically decaying $\tau$-leptons with the ATLAS experiment}

\AtlasAbstract{%
Cross section measurements of high-Lorentz boosted Higgs boson production in final states with pairs of hadronically decaying  $\tau$-leptons are presented.
The analysis is based on the full Run~2 data sample of proton--proton collisions at a centre-of-mass energy of $\sqrt{s} = 13$ TeV, corresponding to an integrated luminosity of 140~$\mathrm{fb}^{-1}$, and a partial Run~3 data sample at $\sqrt{s} = 13.6$ TeV with an integrated luminosity of 162~$\mathrm{fb}^{-1}$ collected in 2022--2024 with the ATLAS experiment.
The analysis extends the sensitivity of previous $H \to \tau\tau$ measurements to Higgs boson transverse momenta above 300 GeV by reconstructing $\tau$-lepton pairs produced with a small separation in large radius jets.
The standard reconstruction of hadronic $\tau$-leptons has limited efficiency in this phase space, requiring dedicated boosted di-$\tau$ reconstruction and identification techniques, combined with multivariate analysis methods, to extract the signal.
For Higgs boson transverse momenta above 300 GeV, results are presented for both inclusive and differential cross-sections. Within the simplified template cross-section framework, differential cross-section measurements in Higgs boson transverse momentum for gluon--gluon fusion production are provided, together with an inclusive cross-section measurement for vector-boson fusion production.
For the inclusive measurement, the analysis yields an observed (expected) significance of 3.8 (3.3) standard deviations over the background-only hypothesis, providing evidence of Higgs boson production in the boosted regime of the di-$\tau$ final state.
}

\AtlasRefCode{HIGP-2024-49}

\PreprintIdNumber{CERN-EP-2026-165}

\AtlasJournal{JHEP}

\AtlasCoverEgroupAnalysisTeam{atlas--analysis-team@cern.ch}

\hypersetup{pdftitle={ATLAS document},pdfauthor={The ATLAS Collaboration}}

\begin{document}

\maketitle

\tableofcontents

\section{Introduction}
%

The observation of the Higgs boson in 2012 by the ATLAS and CMS collaborations~\cite{HIGG-2012-27, CMS-HIG-12-028} completed the expected particle content of the Standard Model (SM) of electroweak and strong interactions.
Subsequent measurements of its properties using proton--proton ($pp$) collision data at centre-of-mass energies of 7 and 8~\TeV showed consistency with SM expectations~\cite{HIGG-2014-06, CMS-HIG-14-009}.
With the larger data sample collected from $pp$ collisions at $\sqrt{s} = 13$~\TeV during LHC Run~2, and with further data being collected in Run~3, the programme of Higgs boson property measurements has entered a precision era, allowing more detailed studies and access to previously unexplored kinematic regimes~\cite{HIGG-2021-23, CMS-HIG-22-001}.

The $\Htautau$ decay channel plays a central role in this effort, as it probes the largest leptonic Yukawa coupling.
Since the first observation of this decay in 2016~\cite{HIGG-2015-07}, the ATLAS Collaboration has steadily improved its sensitivity to the $\Htautau$ process~\cite{HIGG-2019-09}.
These efforts culminated in some of the most precise vector-boson fusion (VBF) production cross-section measurements~\cite{HIGG-2022-07} for a single decay channel using the simplified template cross-section (STXS) framework~\cite{badger2016leshouches2015physics,deFlorian:2016spz,Berger_2026}.

At the LHC, Higgs bosons are predominantly produced via gluon--gluon fusion ($gg$F), vector-boson fusion, associated production with a vector boson ($VH$), and associated production with a top-quark pair (\ttH).
The $gg$F production mode of the Higgs boson is particularly important to address open questions in particle physics, since it represents more than 85\% of the event rate. Any deviation in the top-quark Yukawa coupling or modifications to the gluon--gluon fusion loop structure would alter the Higgs boson production rate and kinematic distributions.
These effects can become especially pronounced at high Higgs boson transverse momentum (\pt), where new physics contributions would result in deviations from SM expectations~\cite{Schlaffer_2014, Grojean_2014}.

Measuring Higgs boson properties at high momentum (\pTH) is experimentally difficult, as the decay products become increasingly collimated.
This topology, referred to as the boosted regime, requires challenging reconstruction of hadronically decaying $\tau$-leptons ($\tauhad$)~\cite{ATL-PHYS-PUB-2022-044}, since at high transverse momentum the two $\tauhad$ candidates often merge into a single object.
Standard algorithms that are optimised for well-separated decay products become less efficient.
To address this, dedicated techniques for boosted di-$\tauhad$ reconstruction and identification were developed~\cite{HDBS-2019-22}.

This paper presents the first ATLAS measurements of the cross-sections for SM Higgs bosons decaying into a pair of $\tau$-leptons in the highly Lorentz-boosted regime.
The phase space of interest corresponds approximately to $\pTH > 300$~\GeV.
The measurement is performed in the channel where both $\tau$-leptons decay hadronically.
The results are presented in terms of an inclusive and differential $\Htautau$ cross-section measurements as well as in terms of cross-sections within the STXS framework.

Similar measurements were performed in the $\Htautau$ final state~\cite{CMS-HIG-21-017} by the CMS Collaboration, and in the $H\rightarrow bb$ final state~\cite{HIGP-2024-01} by the ATLAS and~\cite{CMS-HIG-21-020} CMS collaborations. Compared to the measurement in Ref.~\cite{HIGP-2024-01}, the results in this paper extend the measurement's lower \pTH\ down from 450~\GeV to 300~\GeV. %

The paper is organised as follows.
\Cref{sec:detector} describes the ATLAS detector.
This is followed in~\cref{sec:samples} by a description of the data samples and Monte Carlo (MC) samples of simulated events employed in the measurement.
\Cref{sec:objrec} details the reconstruction of the physics objects.
The event selection and categorisation is described in \cref{sec:evsel}.
In \cref{sec:bkgest}, the estimation of the background processes is discussed.
Section~\ref{sec:mva_analysis} describes the multivariate analysis methods used to extract and categorise the signal, while Section~\ref{sec:systematics} presents the estimation of systematic uncertainties affecting the measurement.
The details of the signal extraction and the results are reported in \cref{sec:result}.
\Cref{sec:conclusion} summarises the conclusions of this work.


%
\section{ATLAS detector}
\label{sec:detector}

%
\newcommand{\AtlasCoordFootnote}{%
ATLAS uses a right-handed coordinate system with its origin at the nominal interaction point (IP)
in the centre of the detector and the \(z\)-axis along the beam pipe.
The \(x\)-axis points from the IP to the centre of the LHC ring,
and the \(y\)-axis points upwards.
Polar coordinates \((r,\phi)\) are used in the transverse plane,
\(\phi\) being the azimuthal angle around the \(z\)-axis.
The pseudorapidity is defined in terms of the polar angle \(\theta\) as \(\eta = -\ln \tan(\theta/2)\) and is equal to the rapidity
$ y = \frac{1}{2} \ln \left( \frac{E + p_z}{E - p_z} \right) $ in the relativistic limit.
Angular distance is measured in units of \(\Delta R \equiv \sqrt{(\Delta y)^{2} + (\Delta\phi)^{2}}\).}

The ATLAS detector~\cite{PERF-2007-01,GENR-2019-02} at the LHC is a multipurpose particle detector
with a forward--backward symmetric cylindrical geometry and a near \(4\pi\) coverage in
solid angle.\footnote{\AtlasCoordFootnote}
It consists of an inner tracking detector surrounded by a thin superconducting solenoid
providing a \qty{2}{\tesla} axial magnetic field, electromagnetic (EM) and hadronic calorimeters, and a muon spectrometer.
The inner tracking detector covers the pseudorapidity range \(|\eta| < 2.5\).
It consists of silicon pixel, silicon microstrip, and transition radiation tracking detectors.
Lead/liquid-argon (LAr) sampling calorimeters provide EM energy measurements
with high granularity within the region \(|\eta|< 3.2\).
A steel/scintillator-tile hadronic calorimeter covers the central pseudorapidity range (\(|\eta| < 1.7\)).
The endcap and forward regions are instrumented with LAr calorimeters
for EM and hadronic energy measurements up to \(|\eta| = 4.9\).
The muon spectrometer surrounds the calorimeters and is based on
three large superconducting air-core toroidal magnets with eight coils each.
The field integral of the toroids ranges between \num{2.0} and \qty{6.0}{\tesla\metre}
across most of the detector.
The muon spectrometer includes a system of precision tracking chambers up to \(|\eta| = 2.7\) and fast detectors for triggering up to \(|\eta| = 2.4\).
The luminosity is measured mainly by the LUCID--2 detector that is located close to the beampipe.
A two-level trigger system was used to select events~\cite{TRIG-2016-01,TRIG-2022-01}.
The first-level trigger is implemented in hardware and used a subset of the detector information
to accept events at a rate close to \qty{100}{\kHz}.
This is followed by a software-based trigger that
reduced the accepted rate of complete events to \qty{1.25}{\kHz} and \qty{3}{\kHz} on average in Run~2 and Run~3, respectively,
depending on the data-taking conditions.
A software suite~\cite{SOFT-2022-02} is used in data simulation, in the reconstruction
and analysis of real and simulated data, in detector operations, and in the trigger and data acquisition
systems of the experiment.


%
\section{Data and simulated event samples}
\label{sec:samples}

The $pp$ collision data used in this analysis consists of Run~2 and partial Run~3 data samples. The Run~2 data were collected between 2015 and 2018 at a centre-of-mass energy of $\sqrt{s} = 13$~\TeV, and correspond to an integrated luminosity of $140 \pm 1.2~\ifb$~\cite{DAPR-2021-01} after requiring that all detectors were in good operational status~\cite{DAPR-2018-01}.
The partial Run~3 data were collected between 2022 and 2024 at a centre-of-mass energy of $\sqrt{s} = 13.6$~\TeV, and correspond to an integrated luminosity of $162 \pm 3.1~\ifb$~\cite{ATL-DAPR-PUB-2025-001}.

MC simulated events are used to model the \Htautau signal and most of the backgrounds from SM processes.
A summary of the generators used for the simulation is shown in \cref{tab:generators}.
Similar MC samples were also used in Ref.~\cite{HIGG-2022-07} and only a brief description of the main samples is reported in this section. All MC samples were processed through an ATLAS detector simulation program~\cite{SOFT-2010-01} using \GEANT~\cite{Agostinelli:2002hh}.
The same reconstruction techniques were used for the MC samples as for the data.

Higgs boson production via $gg$F was simulated using the \POWHEGBOX[v2]~\cite{Frixione:2007nw,Nason:2004rx,Frixione:2007vw,Alioli:2010xd,Hartanto:2015uka} generator at next-to-next-to-leading-order (NNLO) accuracy in quantum chromodynamics (QCD).
The MC samples are normalised to the next-to-next-to-next-to-leading-order (N${}^3$LO) cross-section in QCD and to electroweak (EW) corrections at next-to-leading order (NLO)~\cite{deFlorian:2016spz,Anastasiou:2016cez,Anastasiou:2015ema,Dulat:2018rbf,Harlander:2009mq,Harlander:2009bw,Harlander:2009my,Pak:2009dg,Actis:2008ug,Actis:2008ts,Bonetti:2018ukf,Bonetti:2018ukf}. For this process, to simulate the Run~2 (Run~3) data the \PDFforLHC[15nnlo] (\PDFforLHC[21nnlo]) set of parton distribution functions (PDF)~\cite{Butterworth:2015oua} was used, while the parton shower (PS) and non-perturbative effects were modelled by \PYTHIA[8.230]~\cite{Sjostrand:2014zea} with parameter values set to those of the \AZNLO~\cite{STDM-2012-23} (A14~\cite{ATL-PHYS-PUB-2014-021}) set of tuned parameters (tune) and the \CTEQ[6L1]~\cite{Pumplin:2002vw} (\NNPDF[2.3lo]~\cite{Ball:2012cx}) PDF set.

Higgs boson production via VBF was simulated using \POWHEGBOX[v2] at NLO accuracy in QCD\@.
It was tuned to match calculations with effects due to finite heavy-quark masses and soft-gluon resummation up to next-to-next-to-leading logarithms (NNLL).
The MC samples are normalised to an approximate NNLO cross-section in QCD with NLO EW corrections~\cite{Ciccolini:2007jr,Ciccolini:2007ec,Bolzoni:2010xr}.
To simulate the Run~2 (Run~3) data the \PDFforLHC[15nlo] (\PDFforLHC[21nlo]) PDF set was used, while \PYTHIA[8.230] with the \AZNLO (A14) tune and the \CTEQ[6L1] (\NNPDF[2.3lo]) PDF set was used for PS simulation.

Higgs boson production via $VH$ was simulated with \POWHEGBOX[v2] at NLO accuracy in QCD for $VH$ plus one jet production.
The loop-induced $gg\to ZH$ process was generated separately at leading order (LO) in QCD\@.
The MC samples are normalised to cross-sections calculated at NNLO in QCD with NLO EW corrections for $p p \to VH$ and at NLO and next-to-leading-logarithm accuracy in QCD for $gg \to ZH$~\cite{Ciccolini:2003jy,Brein:2003wg,Brein:2011vx,Altenkamp:2012sx,Denner:2014cla,Brein:2012ne,Harlander:2014wda}.
Similarly to the VBF samples, to simulate the Run~2 (Run~3) data the \PDFforLHC[15nlo] (\PDFforLHC[21nlo]) PDF set was used, while \PYTHIA[8.230] with the \AZNLO (A14) tune and the \CTEQ[6L1] (\NNPDF[2.3lo]) PDF set was used for PS simulation.

The production of \ttH events was simulated at NLO accuracy in QCD using \POWHEGBOX[v2], with the decays of bottom and charm hadrons performed by \EVTGEN[1.6.0]~\cite{Lange:2001uf}.
The cross-section used to normalise the \ttH process is calculated at NLO in QCD and EW couplings~\cite{deFlorian:2016spz,Beenakker:2002nc,Dawson:2003zu,Yu:2014cka,Frixione:2014qaa}.
For this sample, to simulate the Run~2 (Run~3) data the \NNPDF[3.0nnlo] (\PDFforLHC[21nlo]) PDF set~\cite{Ball:2014uwa} was used, with the PS modelled by \PYTHIA[8.230] with the A14 tune and the \NNPDF[2.3lo] PDF set.

The normalisation of all Higgs boson samples accounts for the decay branching ratio calculated with \textsc{Hdecay}~\cite{Djouadi:1997yw,Spira:1997dg,Djouadi:2006bz} and \PROPHECY~\cite{Bredenstein:2006ha,Bredenstein:2006rh,Bredenstein:2006nkV2}.
The decays and spin correlations for $\tau$-leptons were handled by \PYTHIA.
The SM prediction of the \Htautau branching ratio ($0.0625 \pm 0.0014$) and a Higgs boson mass of \SI{125.09}{\GeV} are assumed throughout this measurement.

The QCD production of the \Vjets background was simulated with the \SHERPA[2.2.11]/\SHERPA[2.2.14]~\cite{Bothmann:2019yzt} generator using NLO matrix elements (ME) for up to two partons, and LO MEs for up to four partons, calculated with the \textsc{Comix}~\cite{Gleisberg:2008fv} and \OPENLOOPS~\cite{Buccioni:2019sur,Cascioli:2011va,Denner:2016kdg} libraries.
They were matched to the \SHERPA parton shower~\cite{Schumann:2007mg} using the MEPS@NLO prescription~\cite{Hoeche:2011fd,Hoeche:2012yf,Catani:2001cc,Hoeche:2009rj}.
The \NNPDF[3.0nnlo] set of PDFs~\cite{Ball:2014uwa} was used for both ME and PS, and the samples are normalised to an NNLO prediction~\cite{Anastasiou:2003ds}.
The simulated \Vjets events mostly consist of a Z boson recoiling against an energetic jet, while the $W+$jets process contributes through W-boson hadronic decays misidentified as a di-$\tauhad$ object, for which the estimation is based on a data-driven method.
Electroweak production of the $lljj$ final state (with $l=e, \mu,\tau$, and $j$ denoting a jet) was simulated with \SHERPA[2.2.11]/\SHERPA[2.2.14] using LO MEs with up to two additional parton emissions.
For the matching to the PS, the same prescription as for the QCD \Vjets processes was used.
The samples were produced using the VBF approximation~\cite{Jager:2012xk}, which avoids overlap with semileptonic diboson topologies by requiring a $t$-channel colour-singlet exchange.

The production of \ttbar\ events was simulated at NLO in QCD by \POWHEGBOX[v2] with the \NNPDF[3.0nlo] PDF set and with the \hdamp parameter\footnote{The \hdamp\ parameter is a resummation damping factor and one of the parameters that controls the matching of \POWHEG ME to the PS and thus effectively regulates the high-\pt radiation against which the \ttbar\ system recoils.} set to 1.5\,\mtop~\cite{ATL-PHYS-PUB-2016-020}.
The events were interfaced to \PYTHIA[8.230] to model the PS, hadronisation, and underlying event, with parameters set according to the A14 tune and using the \NNPDF[2.3lo] set of PDFs.
The decays of bottom and charm hadrons were performed by \EVTGEN[1.6.0], as they were for the \ttH sample.
The \ttbar sample is normalised to the cross-section prediction at NNLO in QCD, including the resummation of NNLL soft-gluon terms, calculated using \TOPpp[2.0]~\cite{Beneke:2011mq,Cacciari:2011hy,Baernreuther:2012ws,Czakon:2012zr,Czakon:2012pz,Czakon:2013goa,Czakon:2011xx}.

The effect of multiple interactions in the same and neighbouring bunch crossings (\pileup) was modelled by overlaying~\cite{SIMU-2020-01} the simulated hard-scattering event with inelastic $pp$ events generated using both \EPOS[2.0.1.4]~\cite{Werner:2005jf} and \PYTHIA[8.308]~\cite{Bierlich:2022pfr}.
The \EPOS events were generated with the \EPOS LHC tune~\cite{Pierog:2013ria} and the \PYTHIA events with the A3 tune~\cite{ATL-PHYS-PUB-2016-017} and the \NNPDF[2.3lo]  set of PDFs.
\PYTHIA pile-up events include either a high \pt jet, a prompt photon, or a lepton from a $b$-hadron decay, while \EPOS was filtered to simulate all remaining pile-up events in the overlay sample.
The individual simulations were first reweighted to ensure a smooth connection across jet \pt then the combination was reweighted to match the distribution of the actual number of interactions per bunch crossing measured in data.

\begin{table}[t]
\caption{Summary of the MC generators used for the main signal and background samples. Information about the matrix element (ME) generator, parton shower (PS), parton distribution function (PDF) set used in the ME and the PS, and set of tuned parameters (Tune) is provided. The column labelled `Normalisation' specifies the order of the cross-section calculation used for the normalisation of the MC samples.}
\label{tab:generators}
\resizebox{\textwidth}{!}{
\begin{tabular}{l cc cc r r}
\hline \hline
Process               & \multicolumn{2}{c}{Generator}                                   & \multicolumn{2}{c}{PDF set}                                      & Tune  & Normalisation        \\
& ME                                & PS                          & ME                                        & PS                   &       &              \\
\hline \hline
Higgs boson (Run 2)    &                                   &                             &                                           &                      &                      \\
\midrule
$gg$F                   & \POWHEGBOX[v2]                           & \PYTHIA[8]                    & \PDFforLHC[15nnlo]                & \CTEQ[6L1]              & AZNLO & N$^{3}$LO QCD + NLO EW    \\
VBF                   & \POWHEGBOX[v2]                           & \PYTHIA[8]                    & \PDFforLHC[15nlo]                 & \CTEQ[6L1]              & AZNLO & NNLO QCD + NLO EW        \\
$VH$                  & \POWHEGBOX[v2]                           & \PYTHIA[8]                    & \PDFforLHC[15nlo]                 & \CTEQ[6L1]              & AZNLO & NNLO QCD + NLO EW          \\
\ttH                  & \POWHEGBOX[v2]                        & \PYTHIA[8]                    & \NNPDF[3.0nnlo] & \NNPDF[2.3lo]              & A14                  & NLO QCD + NLO EW                  \\
\midrule
Higgs boson (Run 3)    &                                   &                             &                                           &                      &                      \\
\midrule
$gg$F                   & \POWHEGBOX[v2]                           & \PYTHIA[8]                    & \PDFforLHC[21nnlo]                & \NNPDF[2.3lo]              & A14 & N$^{3}$LO QCD + NLO EW    \\
VBF                   & \POWHEGBOX[v2]                           & \PYTHIA[8]                    & \PDFforLHC[21nlo]                         & \NNPDF[2.3lo]              & A14 & NNLO QCD + NLO EW        \\
$VH$                  & \POWHEGBOX[v2]                           & \PYTHIA[8]                    & \PDFforLHC[21nlo]                         & \NNPDF[2.3lo]              & A14 & NNLO QCD + NLO EW          \\
\ttH                  & \POWHEGBOX[v2]                        & \PYTHIA[8]                    & \PDFforLHC[21nlo]  & \NNPDF[2.3lo]              & A14  & NLO QCD + NLO EW                  \\

\midrule
Background            &                                   &                             &                                           &                      &                      \\
\midrule
\Vjets (QCD/EW)      & \multicolumn{2}{c}{\SHERPA[2.2.11/2.2.14]} & \multicolumn{2}{c}{\NNPDF[3.0nnlo]} & \SHERPA                                   & NNLO for QCD, LO for EW                                       \\
\ttbar                & \POWHEGBOX[v2]                           & \PYTHIA[8]                    & \NNPDF[3.0nlo] & \NNPDF[2.3lo]               & A14                  & NNLO + NNLL QCD        \\
\hline \hline
\end{tabular}
}
\end{table}


%
\section{Object reconstruction}
\label{sec:objrec}

Electrons, muons, isolated $\tau$-leptons and small-radius jets are reconstructed and selected following the criteria used in the previous $\Htautau$ coupling measurements described in Ref.~\cite{HIGG-2022-07}.

In particular, small-radius jets are reconstructed by applying the \antikt algorithm~\cite{Cacciari:2008gp,Fastjet}, with a radius parameter of $R=0.4$, to
topological calorimeter-energy clusters and charged-particle tracks processed with a particle-flow algorithm~\cite{PERF-2015-09}.
The energy of the jets is corrected for the non-compensating calorimeter response, noise threshold effects, energy loss from inactive material, and pile-up contamination~\cite{JETM-2018-05}.
Only jets with $\pT>\SI{20}{\GeV}$ and $|\eta|<4.5$ are considered.
A jet vertex tagger (JVT)~\cite{PERF-2014-03} is used to remove jets with $\pT<\SI{60}{\GeV}$ and $\abseta<2.5$ that are identified as not being associated with the primary vertex \footnote{The primary vertex is defined as the reconstructed vertex with the highest sum of squared transverse momenta of the associated tracks~\cite{PERF-2015-01}.} of the event.
Similarly, pile-up jets in the forward region are suppressed with a \enquote{forward JVT}~\cite{PERF-2016-06} algorithm that exploits jet shapes and topological jet correlations in pile-up interactions and is applied to all jets with $\pT<\SI{60}{\GeV}$ and $|\eta|>2.5$.
Jets with $\pT>\SI{20}{\GeV}$ and $|\eta|<2.5$ containing $b$-hadrons ($b$-jets)  are identified using the GN2 $b$-tagging algorithm~\cite{FTAG-2023-05}.

The removal of reconstructed objects stemming from the same energy clusters in the calorimeters (overlap removal) and the computation of the missing transverse energy ($\etmiss$) are also performed following the criteria in Ref.~\cite{HIGG-2022-07}.
Events with any electrons, muons or $b$-jets remaining after the overlap removal are rejected.

Since the analysis targets Higgs bosons produced at high $\pTH$, a dedicated boosted di-$\tauhad$ reconstruction and identification procedure is used to select the Higgs boson decay products.
The collection of visible $\tauhad$ decay products are the neutral and charged hadrons stemming from the $\tau$-lepton decay and are referred to as $\tauhadvis$. The $\tauhadvis$ four-momentum is defined as the sum of the four-momenta of the visible $\tauhad$ decay products.
For a boosted pair of $\tauhad$ with $\dRtauhadvistauhadvis < 0.4$, the standard reconstruction using \antikt seeded jets with a radius parameter $R=0.4$  tends to merge the two $\tauhad$ into one single jet.
Furthermore the standard $\tauhad$ identification efficiency degrades at such small distances, since the track and energy depositions of one $\tauhad$ decay fall into the isolation annulus ($0.2<\Delta R<0.4$) of another $\tauhad$ candidate.
Therefore, a dedicated di-$\tauhad$ reconstruction algorithm was developed within ATLAS and used for the first time in Ref.~\cite{HDBS-2019-22}.
This method extends the sensitivity for hadronically decaying $\tau$-lepton pairs with very small opening angles by first reconstructing a di-$\tauhad$ object in a large \antikt ($R=1.0$) calorimeter seeded jet (LRJ).
The individual $\tauhad$ decay products are then reconstructed using smaller \antikt ($R=0.2$) calorimeter jets, called \emph{subjets}, within the di-$\tauhad$ object.
The subjets are required to have $\pT>\SI{10}{\GeV}$ and $|\eta|<2.5$.
A schematic of the di-$\tauhad$ reconstruction is shown in \cref{fig:omni_diagram}.

\begin{figure}[h!]
\begin{center}
\includegraphics[width=1\linewidth]{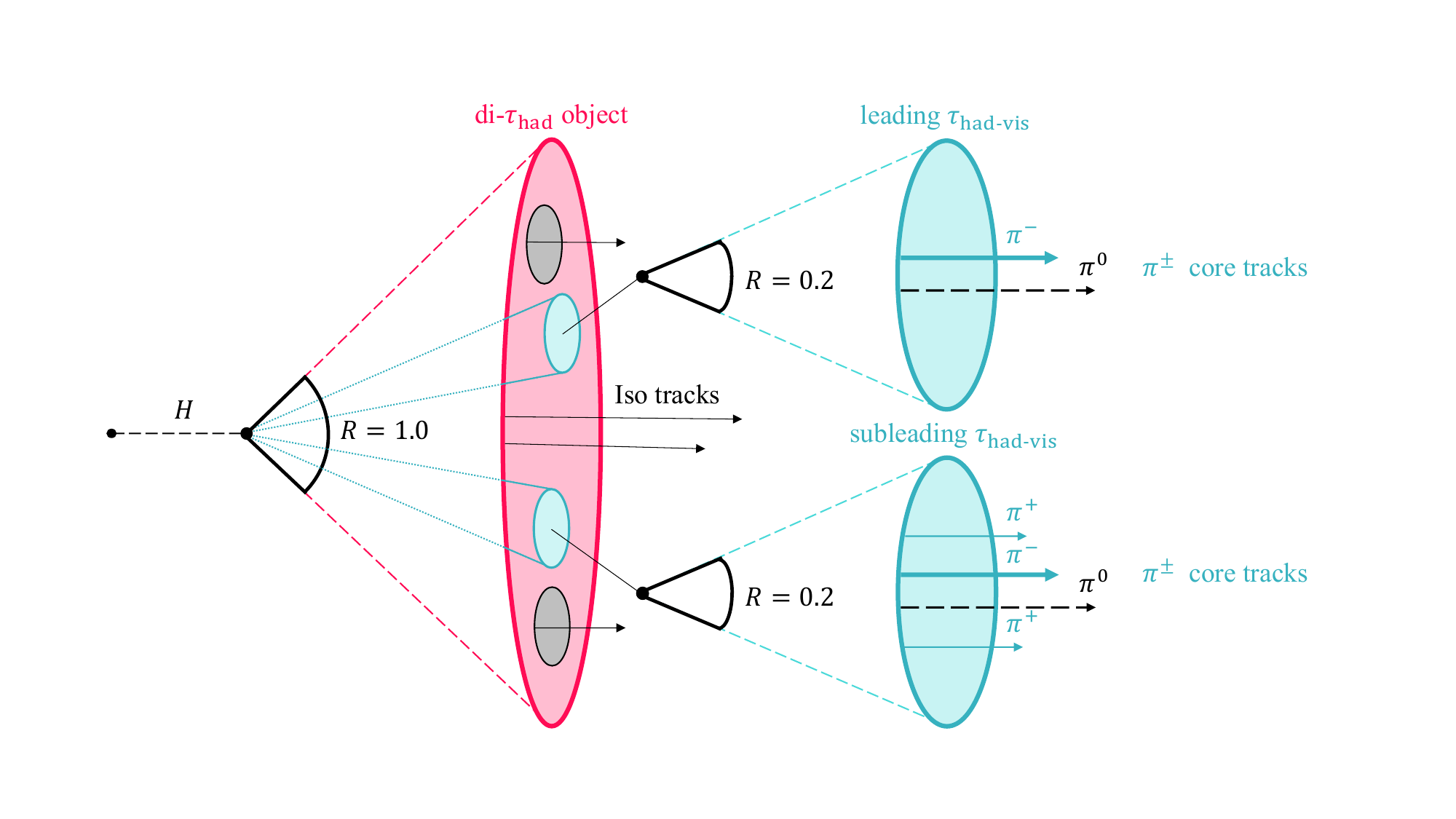} %
\end{center}
\caption{A schematic of the di-$\tauhad$ reconstruction. Starting from a large \antikt ($R=1.0$) calorimeter seeded jet, the individual $\tauhad$ decay products (\emph{subjets}) are reconstructed using smaller \antikt ($R=0.2$) calorimeter jets. The `core-tracks' (`iso-tracks')  correspond to tracks that satisfy the quality requirements and are within a (not within any) di-$\tauhad$ subjet.}
\label{fig:omni_diagram}
\end{figure}

A reconstructed di-$\tauhad$ candidate is required to have $p_{\text{T}}(\mathrm{LRJ}) > 300~\GeV$, $|\eta|<2.5$, and at least two subjets.
The two highest-\pT subjets are expected to contain the decay products of the two boosted $\tauhad$.
The reconstructed di-$\tauhad$ four-momentum and the di-$\tauhad$ direction are calculated from the combined four-momenta of all subjets within the reconstructed di-$\tauhad$ candidate.

The reconstruction efficiency is shown in \cref{fig:reco_eff_ggh} for the standard single-$\tauhad$ and the boosted di-$\tauhad$ reconstruction using a MC sample of Higgs bosons produced via $gg$F and VBF, and subsequently decaying into $\tau$-leptons.
A di-$\tauhad$ candidate is considered to be correctly reconstructed when each of the two leading subjets match the generated \tauhadvis within a cone of $\Delta R < 0.2$.
The efficiency is defined as the number of events where both $\tauhad$ candidates are correctly reconstructed divided by the total number of simulated events in which both $\tau$-leptons decay hadronically, without any condition on the number of reconstructed tracks.
\cref{fig:reco_eff_ggh_dR} shows the efficiency of the standard single-$\tauhad$ reconstruction of $\tauhad$ pairs as a function of the distance between the simulated visible $\tauhad$ decay products, $\Delta R^{\text{gen}}_{\tauhadvis\tauhadvis}$. The efficiency has a threshold starting at $\Delta R^{\text{gen}}_{\tauhadvis\tauhadvis} = 0.4$ with little efficiency below.
Below $\Delta R^{\text{gen}}_{\tauhadvis\tauhadvis} = 0.4$, the two \tauhadvis are merged into a single $R = 0.4$ seeded jet and not reconstructed separately, clearly demonstrating the necessity for an explicit di-$\tauhad$ reconstruction in this regime.
The di-$\tauhad$ reconstruction achieves high efficiency for $\Delta R^{\text{gen}}_{\tauhadvis\tauhadvis}$  down to 0.2, and maintains high efficiency across a broad \pt range starting from the visible generated Higgs boson, $p^{H}_{\text{T, gen--vis}} > 300$~\GeV, defined as the sum of the four-momenta of the two generated \tauhadvis from the Higgs boson decay.

\begin{figure}[h!]
\begin{center}
\subfloat[]{\includegraphics[width=0.475\textwidth]{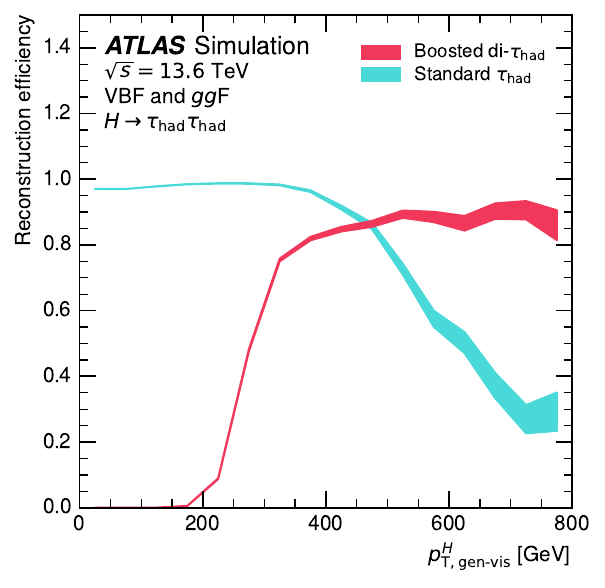}\label{fig:reco_eff_ggh_pthvis}}
\subfloat[]{\includegraphics[width=0.475\textwidth]{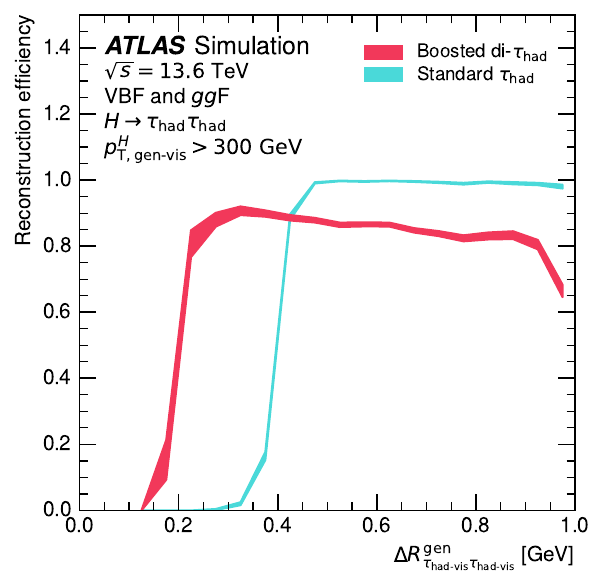}\label{fig:reco_eff_ggh_dR}}
\caption{Di-$\tauhad$ reconstruction efficiency for the standard single-$\tauhad$ reconstruction and the di-$\tauhad$ reconstruction as a function of (a) the visible generated Higgs boson $\pt$, $p^{H}_{\text{T, gen--vis}}$,  and (b) the distance $\Delta R^{\text{gen}}_{\tauhadvis\tauhadvis}$ between the simulated visible $\tauhad$ decay products for signal events. For (b), an additional requirement of $p^{H}_{\text{T, gen--vis}} > 300$~\GeV is applied to highlight the boosted phase space regime of the analysis. The solid bands include the statistical uncertainties due to the number of simulated events.}
\label{fig:reco_eff_ggh}
\end{center}
\end{figure}

While the reconstruction algorithm is focused on high acceptance of di-$\tauhad$ candidates, the identification algorithm, referred to as `OMNI tagger', is optimised to separate real di-$\tauhad$ decays from the large amount of multi-jet background.
Advanced machine learning architectures capable of leveraging low-level track information were explored.
The point-edge transformer (PET) architecture is selected because of its success in jet classification tasks in high-energy physics~\cite{Mikuni_2025}.
PET integrates graph neural network (GNN) components with transformer mechanisms, enabling both local geometric reasoning and global context modelling.
A set of subjet and track variables are used as input features for the OMNI tagger, and are summarised in \cref{tab:jet_features}.
The algorithm training is performed using simulated multi-jet events as background and a sample of Higgs bosons produced via $VH$ with a flat Higgs boson mass distribution as signal. This flat Higgs boson mass distribution of $VH$ events avoids some specific signal features of the target space of the present analysis, allowing the tagger to be used in different phase spaces of other analyses. %

\begin{table}[h!]
\renewcommand{\arraystretch}{1.5}
\centering
\footnotesize
\caption{Variables used in the OMNI tagger training for di-$\tauhad$ identification. LRJ is the seed large-radius jet containing the di-$\tauhad$ object, $\mathrm{sj_1}$ and $\mathrm{sj_2}$ are the first and second subjets ordered in $\pt$, respectively. The `iso-tracks' correspond to tracks that satisfy the quality requirements, but are not within any di-$\tauhad$ subjet. }
\label{tab:jet_features}
\begin{tabular}{p{0.35\linewidth} p{0.6\linewidth}}
\hline \hline
Variable & \mline{Definition} \\
\hline \hline{$E_{\Delta R<0.1}^\mathrm{sj_1}/E_{\Delta R<0.2}^\mathrm{sj_1}$ and $E_{\Delta R<0.1}^\mathrm{sj_2}/E_{\Delta R<0.2}^\mathrm{sj_2}$} & \mline{Ratios of the energy deposited in the core to that in the full cone, for the subjets $\mathrm{sj_1}$ and $\mathrm{sj_2}$, respectively}                                                                           \\
{$\pt^\mathrm{sj_2}/\pt^\mathrm{LRJ}$ and $(\pt^\mathrm{sj_1} + \pt^\mathrm{sj_2})/\pt^\mathrm{LRJ}$}                                 & \mline{Ratio of the $\pt$ of $\mathrm{sj_2}$ to the di-$\tauhad$ large-radius jet $\pt$ and ratio of the scalar $\pt$ sum of the two leading subjets to the di-$\tauhad$ large-radius jet $\pt$, respectively}           \\
{$\log(\sum \pt^{\text{iso-track}}/\pt^\mathrm{LRJ})$}                                                                               & \mline{Logarithm of the ratio of the scalar $\pt$ sum of the iso-tracks to the di-$\tauhad$ large-radius jet $\pt$}                                                                                                      \\
{$\Delta R_{\text{max}}(\text{track}, \mathrm{sj_1})$ and $\Delta R_{\text{max}}(\text{track}, \mathrm{sj_2})$}                       & \mline{Largest separation of a track from its associated subjet axis, for the subjets $\mathrm{sj_1}$ and $\mathrm{sj_2}$, respectively}                                                                                 \\
{$\sum[\pt^{\text{track}}\Delta R(\text{track}, \mathrm{sj_2})]/\sum \pt^{\text{track}}$  }                                          & \mline{$\pt$-weighted sum of $\Delta R$ of the tracks matched to $\mathrm{sj_2}$ relative to its axis}                                                                                                               \\
{$\sum[\pt^{\text{iso-track}}\Delta R(\text{iso-track}, \mathrm{sj})]/\sum \pt^{\text{iso-track}}$}                                  & \mline{$\pt$-weighted sum of $\Delta R$ between iso-tracks and the nearest subjet axis}                                                                                                                                  \\
{$\log(m_{\Delta R<0.1}^{\text{tracks}, \mathrm{sj_1}})$ and $\log(m_{\Delta R<0.1}^{\text{tracks}, \mathrm{sj_2}})$}                 & \mline{Logarithms of the invariant mass of the tracks in the core of $\mathrm{sj_1}$ and $\mathrm{sj_2}$, respectively}                                                                                                  \\
{$\log(m_{\Delta R<0.2}^{\text{tracks}, \mathrm{sj_1}})$ and $\log(m_{\Delta R<0.2}^{\text{tracks}, \mathrm{sj_2}})$}                 & \mline{Logarithms of the invariant mass of the tracks with $\Delta R < 0.2$ from the axis of $\mathrm{sj_1}$ and $\mathrm{sj_2}$, respectively}                                                                          \\
{$\log(|d_0^{\mathrm{sj_1},\text{lead-track}}|)$ and $\log(|d_0^{\mathrm{sj_2},\text{lead-track}}|)$}                                 & \mline{Logarithms of the closest distance in the transverse plane between the primary vertex and the leading track of $\mathrm{sj_1}$ and $\mathrm{sj_2}$, respectively}                                                 \\
{$\Delta\eta(\text{LRJ, track}_{i})$}                                                                                                 & \mline{$\Delta\eta$ between the large-radius di-$\tauhad$ seed jet and each track in it}                                                                                                                                 \\
{$\Delta\phi(\text{LRJ, track}_{i})$}                                                                                                 & \mline{$\Delta\phi$ between the large-radius di-$\tauhad$ seed jet and each track in it}                                                                                                                                 \\
{$\Delta R(\text{LRJ, track}_{i})$}                                                                                                   & \mline{$\Delta R$ between the large-radius di-$\tauhad$ seed jet and each track in it}                                                                                                                                   \\
{$\log(\pt^{\text{track}_{i}})$}                                                                                                      & \mline{Logarithms of the $\pt$ of each track}                                                                                                                                                                                   \\
{$\log(\pt^{\text{track}_{i}}/ \pt^\mathrm{LRJ})$}                                                                                    & \mline{Logarithms of the ratio of each track $\pt$ to the large-radius di-$\tauhad$ seed jet $\pt$}                                                                                                                          \\
{$d_{0}^{\text{track}_{i}}$}                                                                                                          & \mline{Impact parameter of each track in the transverse plane~\cite{PERF-2015-08}}                                                                                                                                                           \\
{$z_{0}^{\text{track}_{i}}$}                                                                                                          & \mline{Impact parameter of each track along the $z$ direction~\cite{PERF-2015-08}}                                                                                                                                                             \\
{$N_{\text{Inner pixel}}^{\text{track}_{i}}$}                                                                                         & \mline{Number of inner most pixel layer hits for each track}                                                                                                                                                             \\
{$N_{\text{pixel}}^{\text{track}_{i}}$}                                                                                               & \mline{Number of pixel hits for each track}                                                                                                                                                                              \\
{$N_{\text{SCT}}^{\text{track}_{i}}$}                                                                                                 & \mline{Number of SCT hits for each track}                                                                                                                                                                                \\
{$q^{\text{track}_{i}}$}                                                                                                              & \mline{Charge of each track}                                                                                                                                                                                             \\
\hline \hline
\end{tabular}
\end{table}

The comparative receiver operating characteristic (ROC) curve in Figure~\ref{fig:omni_roc} shows the improved performance of the OMNI tagger relative to an updated version of the boosted decision tree (BDT) used in Ref.~\cite{HDBS-2019-22}.
The signal efficiency of the OMNI tagger is uniform as a function of \pileup, di-$\tauhad$ \pt and mass across a wide range of phase space.
The OMNI tagger achieves over an order of magnitude better background rejection over a wide range of signal efficiency.
The working point at 85\% efficiency is chosen based on optimisation studies of the signal significance for this analysis.
At this point the background rejection is approximately a factor of ten better than the BDT.
The OMNI tagger improvement relative to the BDT is due to the increased number of input features in the training, for example the track related quantities for each track in the subjets, and also to the more complex and performant algorithm architecture.

\begin{figure}[h!]
\begin{center}
\includegraphics[width=0.5\linewidth]{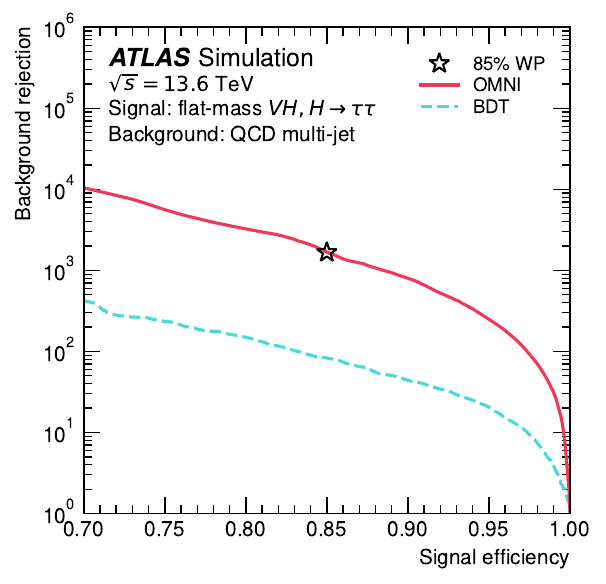}
\end{center}
\caption{ROC curve showing the performance of the OMNI tagger and the BDT. The performance is evaluated using a flat Higgs boson mass $VH$ simulated sample as signal and simulated multi-jet events as background, considering di-$\tauhad$ candidates with $p_{\text{T}}(\mathrm{LRJ}) > 300~\GeV$. The OMNI tagger achieves more than an order of magnitude improvement in background rejection over a wide range of signal efficiency. The $x$-axis shows the signal efficiency, while the $y$-axis shows the background rejection rate. The star symbol denotes the 85\% efficiency working point used in the event selection of the analysis.}
\label{fig:omni_roc}
\end{figure}

A dedicated di-$\tauhad$ energy calibration is also developed as part of the analysis strategy, following the procedure described in Ref.~\cite{PERF-2013-06}. The energy calibration aims to correct the two subjets containing the $\tauhad$ decay products and proceeds in two steps.
In the first step, a MC--MC correction is derived by comparing reconstructed $\tauhad$ candidates in simulation to their corresponding matched generator level $\tau$-lepton decays.
This calibration is derived using a $G\rightarrow HH \rightarrow 4 \tau$ sample, where the graviton mass was set to 1.5~\TeV.
This mass choice provides a large sample of events for the calibration within the kinematic phase space of the analysis.
After having applied the first correction, the second step consists of an in situ correction performed in the final fit, comparing data and simulated events, which provides the final calibration.  %

The MC--MC step for the $\tau$-lepton energy scale (TES) calibration relies only on calorimeter based information.
This calibration largely accounts for the non-compensating response of the calorimeter, as well as energy losses in dead material and outside the clusters used to build the subjets.
The calibrated $\tau$ candidate energy is computed as
\begin{equation}
E_{\text{calib}} = \dfrac{E_{\text{LC}}-E_{\mu}}{\mathcal{R}(E_{\text{LC}}-E_{\mu}, |\eta|, n_{p})}
\end{equation}
where $E_{\text{LC}}$ is the energy of the reconstructed $\tauhad$ candidate at the local hadronic calibration (LC) scale~\cite{PERF-2014-07}, $\mathcal{R}(E_{\text{LC}}, |\eta|, n_{p})$ is the response function, %
and $E_{\mu}$ is a correction to the energy of the reconstructed $\tauhad$ candidate that accounts for the energy deposited by pile-up interactions~\cite{PERF-2013-06}.
The response function is defined as the ratio of the reconstructed $\tau$ candidate  energy (after subtracting the pile-up correction) to the generator-level $\tau$-lepton energy, and is derived as a function of $E_{\text{LC}}$, $|\eta|$ and the number of tracks (prongs), $n_{p}$, in each $\tau$-lepton decay, as shown in Figure~\ref{fig:tes_response_functions}.
This generator-level calibration is applied to both data and simulated events. It is computed separately for one- and three-prong $\tau$-lepton decays in each  considered  $|\eta|$ interval, as shown in \cref{fig:tes_response_functions}. The energy response as a function of $E_{\text{LC}}$ showed a similar behavior as in Ref.~\cite{PERF-2013-06} for the  standard single-$\tauhad$ reconstruction. %

\begin{figure}[h!]
\begin{center}
\begin{tabular}{cc}
\includegraphics[width=0.475\textwidth]{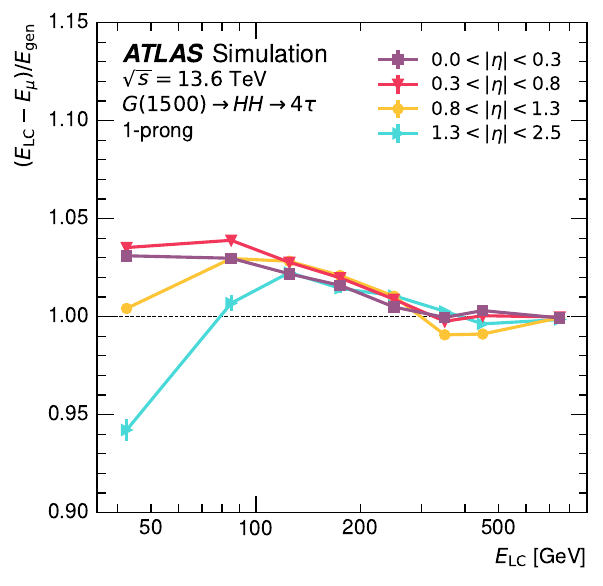} &
\includegraphics[width=0.475\textwidth]{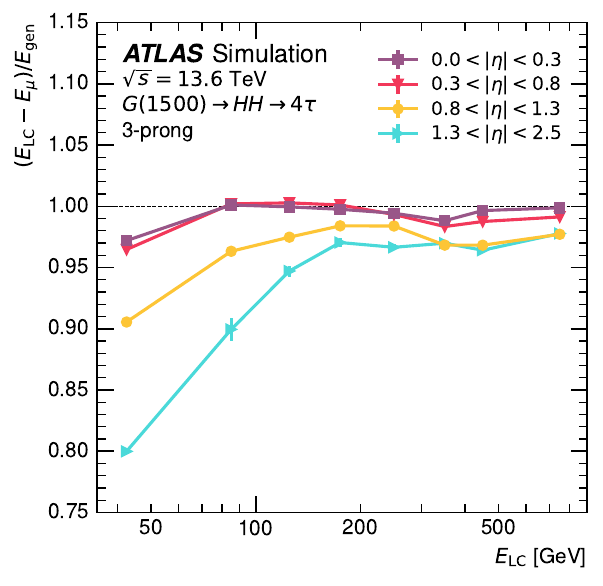} \\
(a) & (b) \\
\end{tabular}
\caption{$\tau$-lepton energy response from simulation for (a) one-prong and (b) three-prong $\tauhad$ decays as a function of the energy of the reconstructed $\tauhadvis$ candidate at the local hadronic calibration (LC) scale, $E_{\text{LC}}$, in different $|\eta|$ intervals. $E_{\mu}$ is a correction to the energy of the reconstructed $\tauhadvis$ candidate that accounts for the energy deposited by pile-up interactions, and $E_{\text{gen}}$ corresponds to the energy of the generator level $\tauhadvis$ candidate. The response is derived using a MC sample of gravitons decaying into a four $\tau$-lepton final state, with the graviton mass set to 1500~\GeV. }
\label{fig:tes_response_functions}
\end{center}
\end{figure}

An initial estimate of the TES correction for the second step is derived from a $\chi^2$ fit for the optimal scale factor of the subjet four-momenta. The fit is performed using the visible mass distribution comparing simulation with data using a $Z \rightarrow \tau \tau$ enriched phase space (with the visible mass less than 80~\GeV), and considers only the statistical uncertainty. The events are selected inclusively in the number of tracks in the leading and subleading di-$\tauhad$ subjets. For the Run 2 (Run 3) data, the $\chi^2$ minimum corresponds to a scaling factor of 1.013 (1.028). The best fit TES estimate is applied to the pre-fit MC templates,  while the final estimation is obtained in the likelihood fit described in \cref{sec:result}.  %

%
%
%
%
%
%


%
\section{Event selection and definition of analysis regions}
\label{sec:evsel}

\newenvironment{DIFnomarkup}{}{}

The trigger and corresponding selection requirements for the analysis are summarised in Table~\ref{tab:trigger_selection} for the Run~2~\cite{TRIG-2016-01} and partial Run~3~\cite{TRIG-2022-01}  data-taking periods.
For each year, the trigger strategy relied on a logical OR of the available single small-$R$ jet, large-$R$ jet, $\tauhad$, and $\etmiss$ triggers, always selecting the lowest unprescaled option.
Events are assured to fall well within the efficiency plateau region of their corresponding trigger by requiring a minimum value for the $\etmiss$ or $\pt$ of the offline reconstructed objects that correspond to the online trigger-level objects.
Overall, the trigger efficiency for signal events, after all other analysis selections are applied, is approximately 98$\%$. %
Due to the high efficiency, no corrections are applied or uncertainties are considered for the trigger efficiency to the simulated events.

\begin{DIFnomarkup}\begin{table}[tb]
\caption{Analysis trigger selection for the Run~2 and Run~3 data. `Online Selection' (`Offline Selection') indicates the requirements for the trigger-level (event-level) reconstructed objects while $\mathrm{sj_1}$ refers to the first di-$\tauhad$ subjet ordered in $\pt$.  }
\label{tab:trigger_selection}
\centering
\footnotesize
\begin{tabular}{llrr}
\hline \hline
Data Period             & Trigger Type     & {Online Selection [\GeV]}       & {Offline Selection [\GeV]}        \\
\hline \hline
\multirow{2}{*}{2015}   & Small-$R$ jet    & {$\pt^\mathrm{jet} > 360$}     & {$\pt^{\text{lead jet}} > 380$}  \\
& $\etmiss$        & {$\etmiss > 70$}               & {$\etmiss > 200$}                \\
\midrule
\multirow{3}{*}{2016}   & Small-$R$ jet    & {$\pt^\mathrm{jet} > 380$}     & {$\pt^{\text{lead jet}} > 400$}  \\
& Large-$R$ jet    & {$\pt^\mathrm{jet} > 420$}     & {$\pt^\mathrm{LRJ} > 470$ }      \\
& $\etmiss$        & {$\etmiss > 100$}              & {$\etmiss > 200$}                \\
\midrule
\multirow{4}{*}{2017}   & Small-$R$ jet    & {$\pt^\mathrm{jet} > 400$}     & {$\pt^{\text{lead jet}} > 420$ } \\
& Large-$R$ jet    & {$\pt^\mathrm{jet} > 440$}     & {$\pt^\mathrm{LRJ} > 490$ }      \\
& $\etmiss$        & {$\etmiss > 110$         }     & {$\etmiss > 200$}                \\
& \tauhad              & {$\pt^{\tauhad} > 160$}     & {$\pt^{\mathrm{sj_1}} > 250$}             \\
\midrule
\multirow{4}{*}{2018}   & Small-$R$ jet    & {$\pt^\mathrm{jet} > 420$}     & {$\pt^{\text{lead jet}} > 440$}  \\
& Large-$R$ jet    & {$\pt^\mathrm{jet} > 420$}     & {$\pt^\mathrm{LRJ} > 470$}       \\
& $\etmiss$        & {$\etmiss > 110$         }     & {$\etmiss > 200$}                \\
& \tauhad              & {$\pt^{\tauhad} > 160$}     & {$\pt^{\mathrm{sj_1}} > 250$}             \\
\midrule
\multirow{4}{*}{2022/2023}  & Small-$R$ jet   & $\pt^{\text{jet}} > 420$  & $\pt^{\text{lead jet}} > 440$ \\
& Large-$R$ jet   & $\pt^{\text{jet}} > 420$  & $\pt^{\text{LRJ}} > 470$ \\
& $\etmiss$       & $\etmiss > 90$            & $\etmiss > 200$ \\
& \tauhad         & $\pt^{\tauhad} > 160 $        & $\pt^\mathrm{sj_1} > 200$ \\
\midrule
\multirow{4}{*}{2024}       & Small-$R$ jet   & $\pt^{\text{jet}} > 400$  & $\pt^{\text{lead jet}} > 420$ \\
& Large-$R$ jet   & $\pt^{\text{jet}} > 460$  & $\pt^{\text{LRJ}} > 510$ \\
& $\etmiss$       & $\etmiss > 90$            & $\etmiss > 200$ \\
& \tauhad         & $\pt^{\tauhad} > 160 $ & $\pt^{\mathrm{sj_1}} > 200$ \\
\hline \hline
\end{tabular}
\end{table}\end{DIFnomarkup}

After the trigger requirements, additional kinematic and quality selections are applied to enhance a boosted topology and suppress background, as shown in Table~\ref{tab:evt_sel}.
For the di-$\tauhad$ selection, the leading and subleading subjets ordered in $\pt$ are required to have one or three tracks.
The selected events that satisfy the OMNI tagger score criteria (ID region) are enriched in di-$\tauhad$-like events (events with similar characteristics to di-$\tauhad$ events), while the events that fail to satisfy the OMNI tagger score criteria (anti-ID region) are dominated by background, in particular multi-jet QCD events.
Additionally, events are divided into opposite-sign (OS) and same-sign (SS) categories based on the subjet charges, with the OS selection targeting the $\Htautau$ signal and the SS selection targeting background dominated regions.
The signal enriched region (SR) corresponds to events with OS subjets that satisfy the OMNI tagger score threshold.
The control region (CR), defined as OS events failing to satisfy the OMNI tagger score, is used to estimate the misidentified $\tauhad$ background contribution in the SR, as described in Section~\ref{sec:bkgest}.

\begin{DIFnomarkup}\begin{table}[tb]
\caption{Summary of event selection requirements. The number of $b$-jets is the number of jets satisfying the GN2~\cite{FTAG-2023-05} tagger working point with 70\% efficiency. The $\etmiss$ requirement is only applied to events not accepted uniquely by the $\etmiss$ trigger. $\mathrm{sj_1}$ and $\mathrm{sj_2}$ refer to the first and second di-$\tauhad$ subjets ordered in $\pt$, while $q_\mathrm{sj_1}$ and $q_\mathrm{sj_2}$ refer to their respective charges. $E_{\mathrm{vis},i}/E_{\tau,i}$ is the fraction of each $\tau$-lepton's momentum carried by the subjet in the collinear approximation~\cite{collinear_approximation}. Requiring this ratio to be positive ensures the $\etmiss$ is compatible with the neutrinos from $\tau$-lepton decay.}
\label{tab:evt_sel}
\centering
\footnotesize
{
\renewcommand{\arraystretch}{1.1}
\begin{tabular}{l c c}
\hline \hline
Object                          & Signal region                                                              & Control region                       \\
\hline \hline
\multirow{4}{*}{Di-$\tauhad$}   & Satisfy 85\% OMNI WP & Fail 85\% OMNI WP                                      \\
& \multicolumn{2}{c}{$N_{\text{subjets}} \geq 2$}                            \\
& \multicolumn{2}{c}{$\Delta R(\text{di-}\tauhad, \text{lead jet}) > 0.4$} \\
& \multicolumn{2}{c}{$N_{b\text{-jets}} = 0$}                                \\

\midrule
\multirow{4}{*}{Subjets}        & \multicolumn{2}{c}{$N_{\text{tracks}}^\mathrm{sj_1,sj_2} \in \{1,3\}$}            \\
& \multicolumn{2}{c}{$\pt^\mathrm{sj_1} > 50$~\GeV}                                  \\
& \multicolumn{2}{c}{$\pt^\mathrm{sj_2} > 15$~\GeV}                                  \\
& \multicolumn{2}{c}{$q_\mathrm{sj_1} * q_\mathrm{sj_2} = -1$}                             \\
\midrule
\multirow{1}{*}{\etmiss}        & \multicolumn{2}{c}{$\etmiss > 60$~\GeV}                                                                        \\
\midrule
\multirow{2}{*}{Combined}   & \multicolumn{2}{c}{$E_{\mathrm{vis},i}/E_{\tau,i} > 0$} \\
& \multicolumn{2}{c}{$N_{\text{tracks}}^{\text{di-}\tauhad} - N_{\text{tracks}}^\mathrm{sj_1} - N_{\text{tracks}}^\mathrm{sj_2} = 0$}   \\
& \multicolumn{2}{c}{$|\Delta \phi(\etmiss, \mathrm{sj_1+sj_2})| < 1.3$}                                                 \\
\hline \hline
\end{tabular}
}
\end{table}\end{DIFnomarkup}


%
\section{Background estimation}
\label{sec:bkgest}

The $\zttplusjets$ production represents the most significant and irreducible background due to its kinematic similarity to the Higgs boson signal and high production rate; the primary distinguishing feature is the mass difference between the $Z$ and Higgs bosons.
The shape of this background is estimated by using MC samples generated as described in Section~\ref{sec:samples}, and for the final extraction of the analysis results the normalisation of this process is treated as a free parameter in the fit.

The misidentified di-$\tauhad$ background, referred to as \emph{fake} di-$\tauhad$ background, predominantly consists of multi-jet QCD interactions and hadronic decays of $W$ bosons produced with a large boost.
These processes can have quark or gluon jets that are misidentified as di-$\tauhad$ decays.
These backgrounds are estimated by using the fake factor method~\cite{TAUP-2023-01}, whose core function consists of deriving a transfer function, referred to as \emph{fake factor}, from specific regions (CR).
This is then used to predict the number of fake di-$\tauhad$  in the SR\@.
The fake factor is defined as the ratio of the number of events in the ID SS region to those in the anti-ID SS region, after subtracting generator-matched di-$\tauhad$ simulated events, mainly from  $\zttplusjets$ with a small contribution from  $W$+jets, from the data in each region to isolate the fake di-$\tauhad$ background component.
The distribution of the misidentified di-$\tauhad$ background in the SR  is obtained by multiplying the number of events observed in the anti-ID OS region by the fake factor.
The fake factors are parameterised as a function of the subleading subjet $\pt$ and the number of tracks within each subjet, since these  variables are found to best capture the fake di-$\tauhad$ background dependency in the analysis phase space.

Other smaller sources of background are modelled using simulation. These include $\zllplusjets$ production (with $\ell \equiv \mu\, \text{or}\, e$), top quark production, $W+$jets production, diboson production and Higgs boson decays into $W$ bosons.

%


%
\section{Analysis categories definition}
\label{sec:mva_analysis}

%
To enhance the signal sensitivity, a multivariate analysis (MVA) approach is adopted.
Two separate MVAs are applied: one for signal and background separation, and another one for event classification for the differential and STXS measurements. The same MVA framework is used in both cases.
The first MVA employs a two-class BDT, implemented using the \texttt{XGBoost} framework~\cite{xgboost}, to classify events into signal and background regions. The two-class BDT is trained using high-level input features of the signal and background samples. The background samples include the $\zttplusjets$ process, the fake di-$\tauhad$ background estimated in the SR and the other smaller sources of background described in Section~\ref{sec:bkgest}.
The complete set of input features is listed in Table~\ref{tab:mva_features}.
These are derived from the four-momenta of the leading and subleading subjets, considered as $\tauhadvis$ candidates, as well as from the missing transverse energy. Example distributions of such features are illustrated in \cref{fig:mva_run_2_panel1} for the invariant mass estimated under the collinear approximation hypothesis, $m_\mathrm{collinear}$, and the angular separation between the leading and subleading subjets, $\Delta R(\mathrm{sj_1},\mathrm{sj_2})$.
To avoid overtraining and ensure unbiased evaluation of the BDT response, a three-fold cross-validation technique is employed.
This method preserves the statistical power of the full data sample while avoiding information leakage between training and evaluation samples.
Two separate discriminants are trained, one for Run~2 and one for Run~3, to account for the differences in the data-taking conditions, trigger configurations, and detector performance between the two runs.
The two discriminants show similar performance and their output score is used as the fit variable to extract the results presented in \cref{sec:result}.

\begin{table}[tb]
\renewcommand{\arraystretch}{1.5}
\centering
\footnotesize
\caption{Set of features used to train the two-class BDT to separate the signal from background. $\mathrm{sj_1}$ and $\mathrm{sj_2}$ refer to the first and second di-$\tauhad$ subjets ordered in $\pt$. }
\label{tab:mva_features}
\begin{tabular}{p{0.2\linewidth} p{0.75\linewidth}}
\hline \hline
Variable & \mline{Definition} \\
\hline \hline{$\pt^\mathrm{sj_1}$}                                          & \mline{$\pt$ of the leading subjet}                                                                                                           \\
{$\pt^\mathrm{sj_2}$}                                          & \mline{$\pt$ of the subleading subjet}                                                                                                        \\
{$m_\mathrm{vis}$}                                             & \mline{Visible mass, defined as the mass of the vectorial sum of the leading and subleading subjets}                                          \\
{$m_\mathrm{collinear}$}                                       & \mline{Invariant mass estimated under the collinear approximation hypothesis}                                                                 \\
{$\Delta R(\mathrm{sj_1},\mathrm{sj_2})$}                      & \mline{Angular separation between the leading and subleading subjets}                                                                         \\
{$p_{\mathrm{T,\, reco}}^{H}$}                                 & \mline{Magnitude of the vectorial sum of two leading di-$\tauhad$ subjets and \etmiss }                                                   \\
{$\etmiss$}                                                    & \mline{Missing transverse energy}                                                                                                             \\
{$\etmiss$ centrality}                                & \mline{Quantification of where the $\etmiss$ lies between the two subjets of the di-$\tauhad$ system in the azimuthal plane }             \\
{$E_\mathrm{vis}^{\mathrm{sj_1}}/E_{\tauhad}^{\mathrm{sj_1}}$} & \mline{Ratio of visible energy to the total energy of the leading subjet}                                                                 \\
{$E_\mathrm{vis}^{\mathrm{sj_2}}/E_{\tauhad}^{\mathrm{sj_2}}$} & \mline{Ratio of visible energy to the total energy of the subleading subjet}                                                              \\
\hline \hline
\end{tabular}
\end{table}

\begin{figure}[h!]
\begin{center}
\begin{tabular}{cc}
\includegraphics[width=0.475\textwidth]{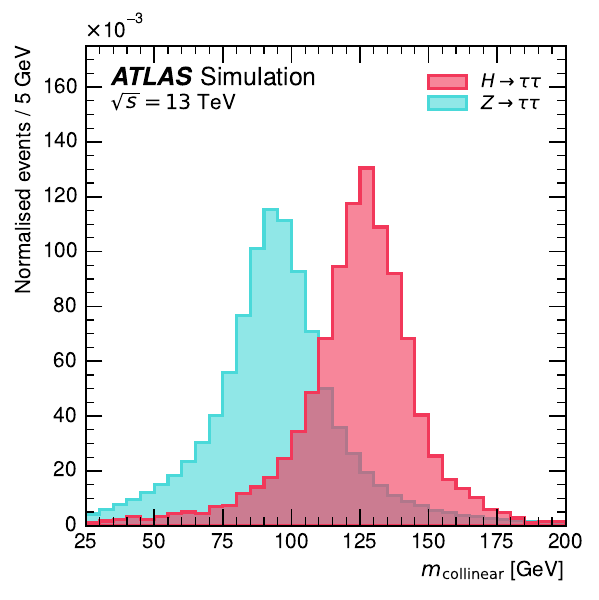} &
\includegraphics[width=0.475\textwidth]{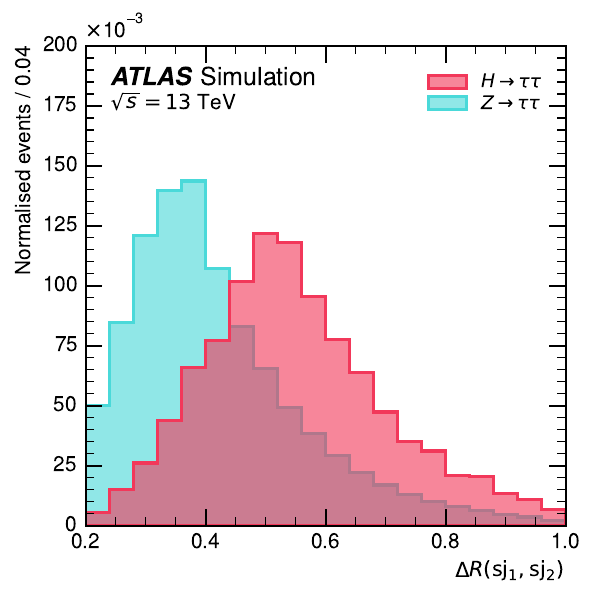} \\
(a) & (b) \\
\end{tabular}
\caption{Higgs boson and $\zttplusjets$ distribution of (a) the collinear mass, $m_{\text{collinear}}$, defined as the invariant mass of the leading and subleading subjets assuming all missing transverse energy is collinear with the visible \tauhad decay products, and (b) the angular separation between the leading and subleading subjets, $\Delta R(\mathrm{sj_1},\mathrm{sj_2})$. The distributions are derived in the signal region defined in \cref{tab:evt_sel}.}  %
\label{fig:mva_run_2_panel1}
\end{center}
\end{figure}

After developing a discriminant between the signal and background processes, a second  MVA-based categorisation is developed to enhance the sensitivity of the Higgs boson measurement in different regions of the phase space being considered.
The analysis measures the Higgs boson cross-section inclusively for $\pTH> 300$~\GeV  as well as four cross-sections in different STXS bins: one for an inclusive VBF production with $\pTH> 300$~\GeV and three for $gg\text{F}$ production in the $\pTH$ bins: 300--450~\GeV, 450--650~\GeV and $> 650$~\GeV. A four-class BDT is used to classify events into the four STXS bins -- the inclusive VBF and three $gg\text{F}$ $\pTH$ bins.
After verifying that the input feature distributions are compatible, the BDT is trained using combined Run~2 and Run~3 simulated events to increase statistical precision of the training samples.
The corresponding set of input features is listed in Table~\ref{tab:stxs_mva_features}.
The variables include a large number of those used in the signal/background two-class classifier training with additional variables targeting both the differences between VBF and $gg\text{F}$ events and between the Higgs boson $\pt$ bins. To build some of these additional variables, $\text{jet}_1$ and $\text{jet}_2$ are defined as the two highest-$\pt$ small-$R$ ($R=0.4$) jets. An explicit requirement is placed on $\text{jet}_1$ to ensure it does not overlap with the di-$\tauhad$ subjets. In the rare case where $\text{jet}_2$ cannot be defined, each variable is set to a default value. Among the variables used, $\Delta \eta(\text{jet}_1,\text{jet}_2)$ has the strongest discriminating power between $gg\text{F}$ and VBF, while $p_{\mathrm{T,\, reco}}^{H}$ and $\Delta R(\mathrm{sj_1},\mathrm{sj_2})$ are dominant in the separation between the STXS $gg\text{F}$ \pTH bins, and are shown in \cref{tab:stxsrank}.

\begin{table}[tb]
\renewcommand{\arraystretch}{1.5}
\caption{Set of features used to train the four-class BDT for STXS categorisation. sj$_1$ and sj$_2$ refer to the first and second di-$\tauhad$ subjets ordered in $\pt$. jet$_1$ and jet$_2$ refer to the two additional small-$R$ ($R=0.4$) jets as defined in the text.}  %
\centering
\footnotesize
\label{tab:stxs_mva_features}
\begin{tabular}{p{0.135\linewidth} p{0.815\linewidth}}
\hline \hline
Variable & \mline{Definition} \\
\hline \hline{$\pt^\mathrm{sj_1}$}                                 & \mline{$\pt$ of the leading subjet }                                                                                                      \\
{$\pt^\mathrm{sj_2}$}                                 & \mline{$\pt$ of the subleading subjet }                                                                                                   \\
{$m_\mathrm{vis}$}                                    & \mline{Visible mass, defined as the mass of the vectorial sum of the leading and subleading subjets }                                     \\
{$m_\mathrm{collinear}$}                              & \mline{Invariant mass estimated under the collinear approximation hypothesis }                                                            \\
{$\Delta R(\mathrm{sj_1},\mathrm{sj_2})$}             & \mline{Angular separation between the leading and subleading subjets }                                                                    \\
{$p_{\mathrm{T,\, reco}}^{H}$}                                              & \mline{Magnitude of the vectorial sum of two leading di-$\tauhad$ subjets and \etmiss }                                                   \\
{$\etmiss$}                                           & \mline{Missing transverse energy }                                                                                                        \\
{$\etmiss\phi$ centrality}                                & \mline{Quantification of where the $\etmiss$ lies between the two subjets of the di-$\tauhad$ system in the azimuthal plane~\cite{HIGG-2013-32} }             \\
{$\Delta R(\mathrm{sj_1},\text{di-}\tauhad)$}         & \mline{Angular separation between the leading subjet and the di-$\tauhad$ system }                                                        \\
{$\eta_{\text{jet}_1}\times\eta_{\text{jet}_2}$}      & \mline{Product of the $\eta$ of the two additional small-$R$ jets }                                                             \\
{$\pt^{\text{jet}_1\text{+}\text{jet}_2}$}            & \mline{Magnitude of the vectorial sum of the two additional small-$R$ jets }                                                    \\
{$p_{\mathrm{T,\, reco}}^{H+jj}$}   & \mline{Magnitude of the vectorial sum of the two leading di-$\tauhad$ subjets, $\etmiss$ and the two additional small-$R$ jets }    \\
{$m_{\text{jet}_1\text{+}\text{jet}_2}$}              & \mline{Invariant mass of the two additional small-$R$ jets }                                                                    \\
{$\Delta R(\text{jet}_1,\text{di-}\tauhad)$}          & \mline{Angular separation between $\text{jet}_1$ and the di-$\tauhad$ system }                                    \\
{$\Delta \eta(\text{jet}_1,\text{jet}_2)$}            & \mline{Distance in $\eta$ between the two additional small-$R$ jets }                                                           \\
{$\pt^{\text{jet}_1}$}                                & \mline{$\pt$ of the highest-$\pt$ additional small-$R$ jet }                                                                                  \\
{$\pt^{\text{jet}_2}$}                                & \mline{$\pt$ of the second highest-$\pt$ additional small-$R$ jet }                                                                           \\
{$\Delta R(\etmiss,\text{jet}_1)$}                    & \mline{Angular distance between $\etmiss$ and $\text{jet}_1$ }                                                    \\
{$\Delta R(\mathrm{sj_1},\text{jet}_1)$}              & \mline{Angular distance between the leading subjet and $\text{jet}_1$  }                                          \\
{$\Delta R(\mathrm{sj_2},\text{jet}_1)$}              & \mline{Angular distance between the subleading subjet and $\text{jet}_1$   }                                       \\
{$\Delta \phi(\etmiss, \mathrm{sj_1\text{+}sj_2})$}   & \mline{Azimuthal angle between the missing transverse energy and the vectorial sum of the leading and subleading subjets }                \\
\hline \hline
\end{tabular}
\end{table}

\begin{figure}[h!]
\begin{center}
\begin{tabular}{cc}
\includegraphics[width=0.475\textwidth]{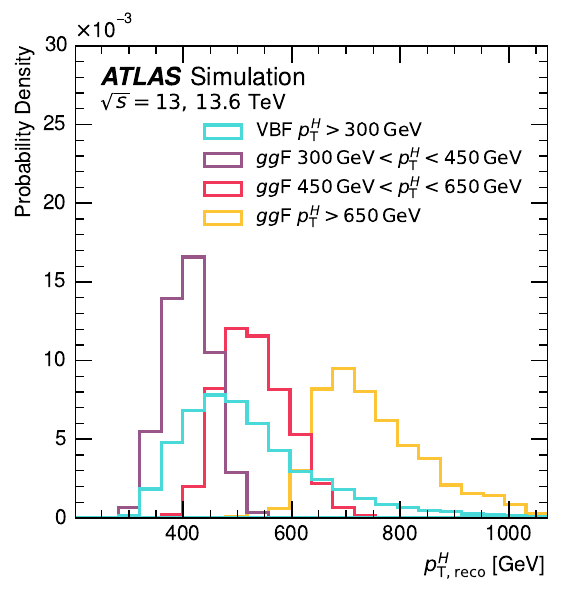} &
\includegraphics[width=0.475\textwidth]{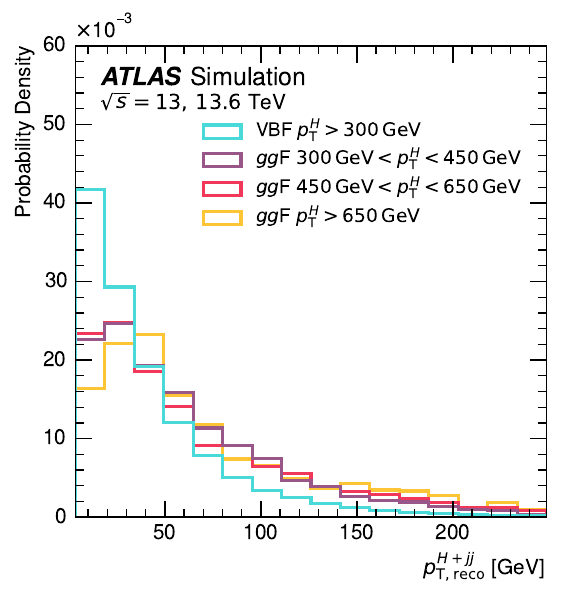} \\
(a) & (b) \\
\includegraphics[width=0.475\textwidth]{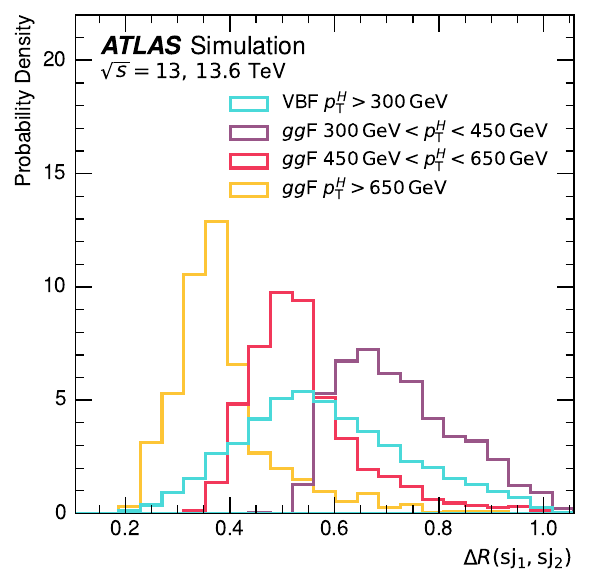} &
\includegraphics[width=0.475\textwidth]{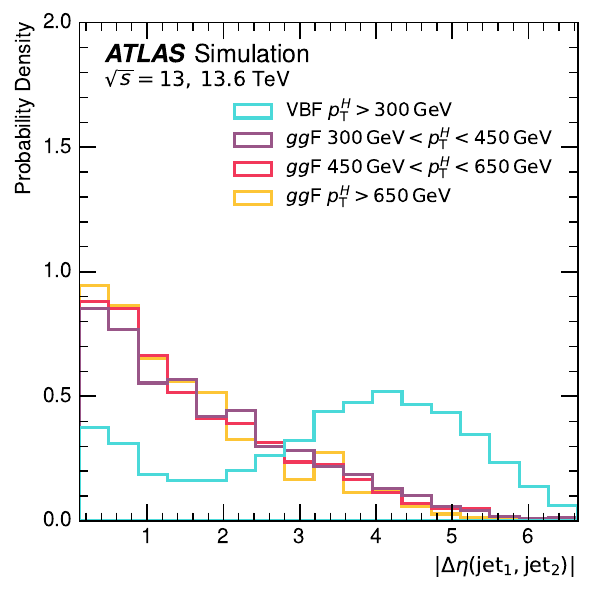} \\
(c) & (d) \\
\end{tabular}
\caption{Probability density distribution of the dominant variables for the four classes of events (with the $gg$F \pt slices defined by the generated Higgs boson \pt):
(a) the magnitude $p_{\mathrm{T,\, reco}}^{H}$ of the vectorial sum of the two leading di-$\tauhad$  subjets and $\etmiss$,
(b) the magnitude $p_{\mathrm{T,\, reco}}^{H+jj}$ of the vectorial sum of the two leading di-$\tauhad$ subjets, $\etmiss$ and the two additional small-$R$ jets,
(c) the angular separation $\Delta{R}(\mathrm{sj_1,sj_2})$ between the leading and subleading subjets of the di-$\tauhad$ candidate, and
(d) the distance $|\Delta\eta(\mathrm{jet_1, jet_2})|$ between the two additional small-$R$ jets.}
\label{tab:stxsrank}
\end{center}
\end{figure}
Each event is classified into the category with the highest score from the four-class BDT\@. The three $gg\text{F}$ enriched regions are referred to as $gg\text{F}(300\ \GeV < \pTH< 450\ \GeV)$, $gg\text{F}(450\ \GeV < \pTH< 650\ \GeV)$, $gg\text{F}(\pTH>650\ \GeV)$.
Within each region, i.e. VBF and the three $gg\text{F}$ $\pTH$ bins, events are again classified into either an SR or $Z\rightarrow\tau\tau$ CR using the two-class BDT output score. The $Z\rightarrow\tau\tau$ CRs from the three $gg\text{F}$ $\pTH$ regions are added together to form the {$gg\text{F-like}$} $Z\rightarrow\tau\tau$ CR.


%
\section{Systematic uncertainties}
\label{sec:systematics}

Systematic uncertainties can affect the normalisations of both signal and background processes, as well as the shape of the discriminating observable used to extract the signal.
They can be separated into three main groups: experimental uncertainties, theoretical uncertainties for the backgrounds and  theoretical uncertainties for the signal.

The experimental systematic uncertainties include those affecting the reconstruction and identification of all final state particles, as well as their energy scale and resolution~\cite{ATLAS:2023dxj,ATLAS-CONF-2017-029,JETM-2018-05,PERF-2014-03,FTAG-2018-01,FTAG-2020-08,FTAG-2019-02}.
The limited size of the MC samples used to build the templates is included among the experimental uncertainties.

The jet energy scale uncertainty includes components related to the in~situ calibration of jets as well as pile-up, the extrapolation to higher transverse momentum, and uncertainties in the different responses to quark- and gluon-initiated jets. For the \etmiss, uncertainties associated with the scale and resolution of the soft-term ~\cite{JETM-2020-03} are considered.

For the di-$\tauhad$ energy calibration, a conservative 100$\%$ uncertainty is assigned on the MC--MC  calibration step, while the final energy scale estimation and the related uncertainty are determined in the likelihood fit described in \cref{sec:result}.
The di-$\tauhad$ energy scaling calibration can induce event migration into the analysis phase space. Considering the requirement of $p_{\text{T}}(\mathrm{LRJ}) > 300~\GeV$ on the di-$\tauhad$ object, applied both in data and simulation, the impact of this migration cannot be determined without bias. To properly quantify the effect, a dedicated $Z\rightarrow\tau\tau$ sample was produced with a $p_{\text{T}}(\mathrm{LRJ}) > 200~\GeV$ requirement and Run~3 detector conditions. The migration induced by the di-$\tauhad$ energy calibration is evaluated using this sample, and the distributions are corrected via a scale factor measured as a function of the vectorial sum of the \pt\ of the leading and subleading subjets. The statistical uncertainty in this correction is taken as an associated uncertainty in the likelihood fit. For Run~2, a 100\% uncertainty is assigned to the correction, as no equivalent sample is available under Run~2 conditions.

An uncertainty associated with the di-$\tauhad$ identification algorithm is derived by studying the modelling in a phase space similar to  the signal region of the analysis.
The OMNI tagger score is examined by selecting events with an 88\% signal efficiency working point. This working point is chosen to minimise the fake di-$\tauhad$ background component. An uncertainty is determined by fitting the ratio of the data to the MC prediction with a linear function. The derived uncertainty is applied to simulated events with a generator matched di-$\tauhad$  for both Run 2 and Run 3. This results in approximately a 5\% uncertainty affecting the signal and background yields. %

The systematic uncertainties for the fake di-$\tauhad$ background include the statistical uncertainty in the fake factors as well as an additional conservative 100\,$\%$ uncertainty on the final background estimated in the SR\@. This latter uncertainty is intended to account for the quark--gluon composition differences between the SR and the SS regions where the fake factors are derived, and also possible differences in the templates of the fake di-$\tauhad$ background between the ID and anti-ID regions. The uncertainty also covers  the largest observed differences when calculating the fake factors using simulated multi-jet events in the OS and SS regions.

For the theoretical uncertainties in the background, the largest impact on the results is due to the uncertainties associated with the \Zjets.
The corresponding uncertainties are considered for the renormalisation scale (\muR), factorisation scale (\muF), and for the choice of $\alphas$ and PDFs.
The effects of uncertainties in \muR and \muF due to missing higher orders are evaluated using six scale variations in the matrix elements by factors of $0.5$ and $2$, excluding the extreme variations in opposite directions~\cite{Cacciari_2004}. %
Several sources of theoretical uncertainty in the signal are considered.
For all production modes, uncertainties are considered for the PDFs and \alphas, the parton shower and hadronisation model, and missing higher orders in the matrix element calculation.
The effects of PDF and \alphas uncertainties are estimated from the \PDFforLHC set of eigenvectors.
The impact of using a different parton shower and hadronisation model is evaluated by comparing the nominal sample with an event sample simulated using \HERWIG[7]~\cite{Bellm:2017jjp} instead of \PYTHIA[8].

For the $gg$F production process, 18 sources of uncertainty related to the $gg$F kinematics and the cross-section are considered, following the prescription in Ref.~\cite{ATL-PHYS-PUB-2023-031}.
Two uncertainties are assigned for the overall fixed-order and resummation effects.
Two uncertainties cover the migrations between different jet-multiplicity bins.
Seven uncertainties are associated with the modelling of the Higgs boson \pt in the different phase space regions.
Four uncertainties account for dijet mass migrations across the STXS bin boundaries.
One uncertainty covers the modelling of the Higgs boson plus two jets transverse momentum distribution in the $\ge$2-jet region.
One uncertainty is assigned to the modelling of the distribution of the Higgs boson plus one jet transverse momentum divided by \pTH in the high-\pTH region. The last uncertainty takes into account the uncertainty from the choice of top-quark mass scheme.

For the VBF and $VH$ processes, uncertainties in the \muR and \muF scale variations are considered. In addition to the uncertainties, the NLO EW corrections calculated for the Higgs boson \pt requirement at \SI{300}{\GeV} are considered. These corrections are derived using the \textsc{Hawk} program~\cite{Denner:2014cla} and no additional uncertainties are assigned to them.

For the \ttH process, six other uncertainties are included: one related to the inclusive cross-section of the process, and five migration uncertainties related to Higgs boson \pt boundaries in the STXS scheme.

The impact of the systematic uncertainties in the cross-section results discussed in \cref{sec:result} is shown in \cref{tab:sig::1poi-breakdown}, where the most impactful are the theoretical uncertainties on the signal and the di-$\tauhad$ related uncertainties.  %

\begin{table}[h!]
\footnotesize
\caption{
Summary of the different sources of uncertainty in decreasing order of their impact on the signal strength in the likelihood fit to measure the cross-section of \Htautau for $\pTH > 300$~\GeV.
Experimental uncertainties for reconstructed objects include efficiency and energy/momentum scale and resolution uncertainties. The uncertainty in the background sample size  includes the bin-by-bin statistical uncertainties in the simulated backgrounds.}
\label{tab:sig::1poi-breakdown}
\begin{center}
\begin{tabular}{lrr}
\hline
\hline
\multirow{2}{*}{Source of uncertainty}      & \multicolumn{2}{r}{Impact on the signal strength} \\
& \multicolumn{1}{r}{Observed} & \multicolumn{1}{r}{Expected}             \\
\hline
Theoretical scale/phase space uncertainty in signal & 0.15  & 0.13  \\
Di-$\tauhad$ energy scale acceptance, Run~2 & 0.12 & 0.08  \\
Misidentified $\tauhad$, Run~3 & 0.10 & 0.08  \\
Background sample size & 0.07  & 0.06  \\
Jet and \etmiss & 0.07  & 0.06   \\
Di-$\tauhad$ identification & 0.04 & 0.03 \\
Parton shower and PDF uncertainty in signal, Run~2 & 0.04 & 0.03 \\
Parton shower and PDF uncertainty in signal, Run~3 & 0.02 & 0.02  \\
Di-$\tauhad$ energy scale acceptance, Run~3 & 0.02 & 0.03 \\
Theoretical uncertainty in \Zjets\ processes & 0.01  & 0.02  \\
Misidentified $\tauhad$, Run~2 & 0.01 & 0.04  \\
Di-$\tauhad$ energy scale MC-MC calibration & 0.01 & 0.01 \\
Luminosity  Run~2 & 0.01  & 0.01   \\
Luminosity  Run~3 & 0.01  & 0.01   \\
\hline
Total systematic uncertainty  & 0.25  & 0.21 \\
\hline
\hline
\end{tabular}
\end{center}
\end{table}


%
\section{Results}
\label{sec:result}

The statistical analysis for the cross-section measurements is based on a likelihood function constructed as the product of Poisson
probability terms over the bins of the input distributions, as shown in Eq.~(\ref{eqn:llh}):

\begin{equation}
\label{eqn:llh}
L\left(\vec{\alpha}, \vec{\theta}, \text{data}\right) = \prod_{k\in\text{cat}}
\prod_{b\in\text{bins}}   P(n_{k,b} | n^{\text{signal}}_{k,b}(\vec{\alpha}, \vec{\theta}) +
n^{\text{bkg}}_{k,b}(\vec{\theta})) \prod_{\theta\in\vec{\theta}} G(\theta),
\end{equation}

where $n_{k,b}$, $n^{\text{signal}}_{k,b}$ and $n^{\text{bkg}}_{k,b}$ refer to the number of observed events, the expected number of signal events and the expected number of background events in the bin $b$ of the analysis region $k$, respectively.
The systematic uncertainties (described in \cref{sec:systematics}) are incorporated in the likelihood function through nuisance parameters, $\vec{\theta}$, that are constrained by Gaussian probability terms, $G$, that multiply the Poisson probability terms, $P$.
The parameters of interest (POIs) of the model are  denoted by  $\vec{\alpha}$ and are estimated by maximising the likelihood.
The test statistic is constructed from the profile likelihood ratio, and the confidence intervals for the POIs are derived using the asymptotic approximation~\cite{Cowan:2010js}.

The likelihood function is built from the following regions:
\begin{itemize}
\item four signal regions, one targeting VBF Higgs boson production and three targeting $gg$F Higgs boson production according to the Higgs boson $\pt$. %
\item two $Z\rightarrow\tau\tau$ control regions, one for VBF-like events and another one for $gg$F-like events. %
\end{itemize}
These regions are defined based on the multivariate analysis described in \cref{sec:mva_analysis}.
The output of the two-class signal BDT is chosen as the fit variable for the signal regions defined by the four-class BDT\@.
For the control regions, the di-$\tauhad$ visible mass is chosen since  it provides better constraints for the TES in situ measurement.

Three different measurements are performed:
\begin{itemize}
\item \emph{Inclusive $pp\to\Htautau$ cross-section for $\pTH > 300$~\GeV}: a single POI fit is performed, considering only the signal events with generator level $\pTH > 300$~\GeV while signal events below this threshold are fixed to the SM expectation. The relative contributions to the $pp\to\Htautau$ cross-section from the various production modes are fixed to the SM predictions. The $\zttplusjets$ normalisation is also determined in the fit through two normalisation factors, one for VBF-like events and another one for $gg$F-like events.
\item \emph{Simplified Template Cross-Section}: four POIs, corresponding to the cross-sections for different bins of the STXS stage 1.2 framework~\cite{Berger_2026}, are determined by the fit. The cross-sections are measured for three $gg$F bins and one EW $qqH$ bin (comprising \text{VBF} and ${qq}\to{V}(\to{qq})H$ production). The same strategy as for the cross-section fit for $\pTH > 300$~\GeV  is used to determine the $\zttplusjets$ normalisation factors.
\item \emph{Differential Cross-Section}: starting from the simplified template cross-section fit set-up, a three POI fit is performed by considering all signal events  independently of the Higgs boson production mode. The signal is measured in three $\pTH$ bins: $300\ \GeV < \pTH<450\ \GeV$, $450\ \GeV < \pTH< 650\ \GeV$ and $\pTH>650\ \GeV$.
\end{itemize}

When combining  Run~2 and Run~3 data samples for the inclusive cross-section measurement, the SM prediction is assumed for the cross-section ratio for the two run periods with different centre-of-mass energies.
\cref{fig:logsoverb} shows a summary of the event yields from the fit to measure the cross-section with $\pTH > 300$~\GeV  as a function of $\log_{10}$(S/B)  for data, background (B), and the Higgs boson signal (S). \cref{fig:postfitrun2} and \cref{fig:postfitrun3} show the signal and control region distributions for the Run~2 and Run~3 data, respectively.
The yield for the fake di-$\tauhad$ background in in the Run 3 dataset is higher than in the Run 2 dataset.
This is due to higher \pileup conditions and the looser trigger strategy for the Run~3 data, allowing more events with fake di-$\tauhad$ candidates to fulfill the analysis selections. %

\cref{fig:1poires} shows the signal strength results for a fit to Run~2, Run~3 and the combined data samples.
The $\zttplusjets$ normalisation values are all compatible with unity within the uncertainties.
The robustness of the $\zttplusjets$ normalisation is tested in an alternative fit, where the $\zttplusjets$ normalisation is split into different \pt bins for the $gg$F region. The results are compatible with the ones quoted in this section.

Results for the significance of the signal relative to a background-only hypothesis are reported in Table~\ref{tab:sgn_values}.
Combining the Run~2 and Run~3 data samples, the analysis reaches an observed (expected) signal significance of 3.8 (3.3) $\sigma$ over the background-only hypothesis, hence providing evidence of Higgs boson production in the boosted regime.
The measured signal strength in the combined fit to the Run~2 and Run~3 data is $\mu_{\pTH>300} = 1.16^{+0.42}_{-0.25} \text{ (Stat.) }^{+0.23}_{-0.27} \text{ (Syst.)}$, and it is compatible at 69\% level with the SM expectation.
The Run~2 and Run~3 results are compatible at 16\% level.
The measured cross-section and the related total uncertainty for the Run~2 (Run~3) data is $\sigma_{\Htautau}^{\pTH>300} = 9.0^{+7.0}_{-4.8} \text{ (Stat.) }^{+2.7}_{-3.3} \text{ (Syst.)}$ (25.2$^{+9.1}_{-6.3}\text{ (Stat.) } ^{5.5}_{-6.2} \text{ (Syst.)}$) fb, in good agreement with the SM prediction of 14$\pm$3 (15.6$\pm$3.5) fb.

The post-fit value of the di-$\tauhad$ energy scale is found to be compatible within one $\sigma$ of the pre-fit value determined by the in situ pre-fit correction described in \cref{sec:objrec}.

\begin{figure}[h!]
\begin{center}
\begin{tabular}{cc}
\includegraphics[scale=0.65]{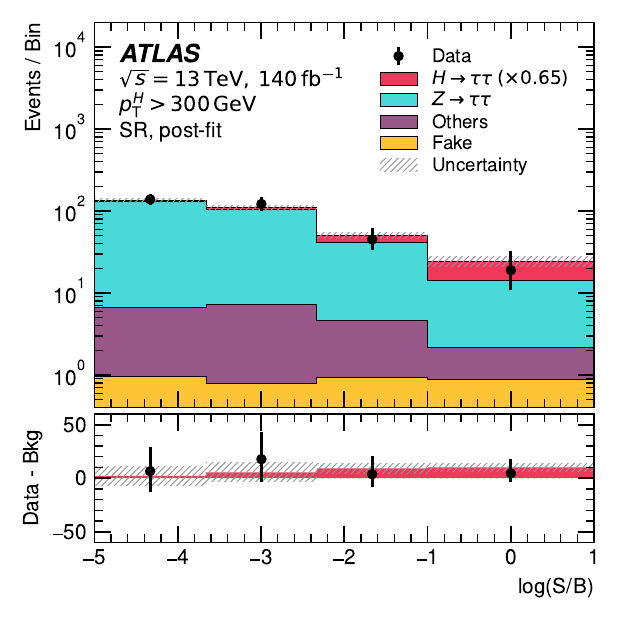} &
\includegraphics[scale=0.65]{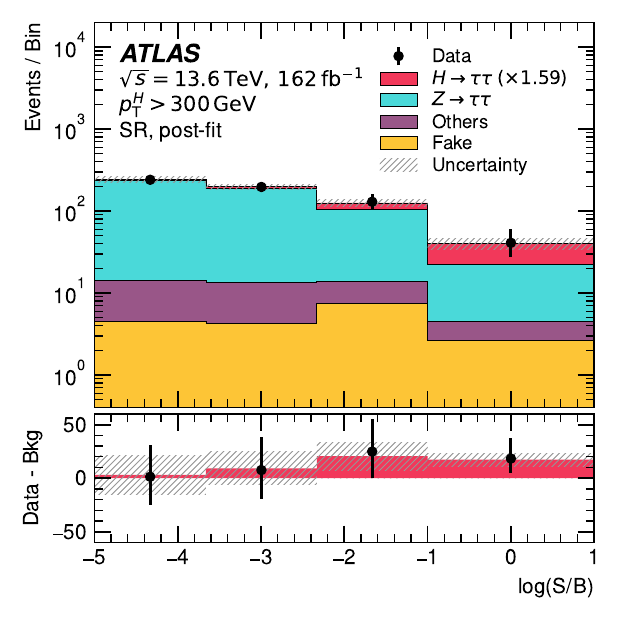} \\
(a) & (b) \\
\end{tabular}
\caption{Event yields from the fit to measure the cross-section of \Htautau with $\pTH > 300$~\GeV as a function of
$\log_{10}$(S/B) for data, background, and the Higgs boson signal for the (a) Run~2 and (b) Run~3 data. The signal is normalised by the signal strength value determined in the fit for each data sample. The $\zttplusjets$ contribution is normalised by two factors, one for VBF-like events and another one for $gg$F-like events. `Others' refer to all minor backgrounds estimated with simulation. The uncertainty bands include both statistical and systematic uncertainties from the fit used to extract the results. The lower panels show the differences between data and background with the solid bands representing the signal contribution. %
}
\label{fig:logsoverb}
\end{center}
\end{figure}

\begin{figure}[h!]
\begin{center}
\begin{tabular}{cc}
\includegraphics[scale=0.54]{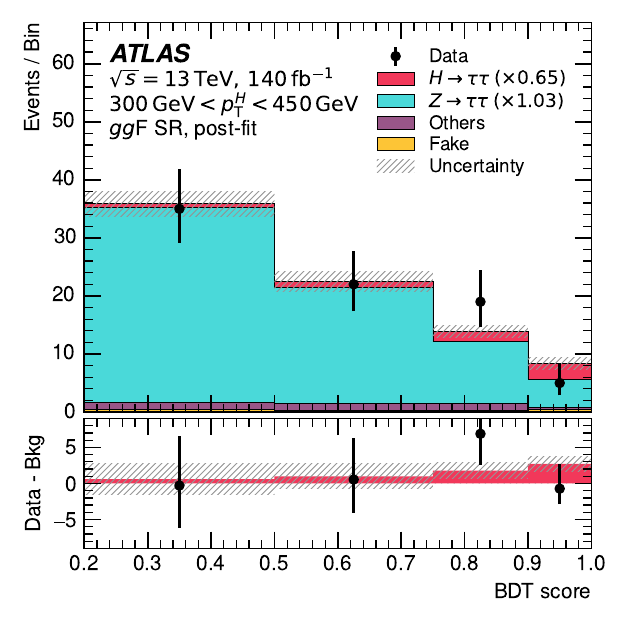} &
\includegraphics[scale=0.54]{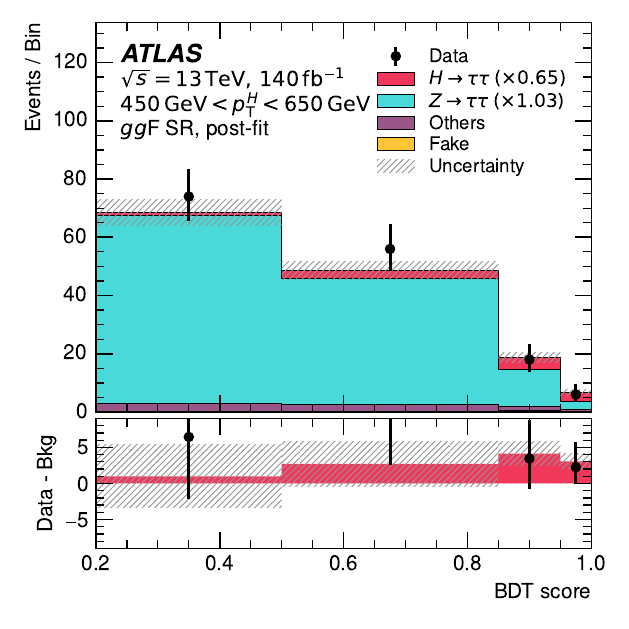} \\
(a) & (b) \\
\includegraphics[scale=0.54]{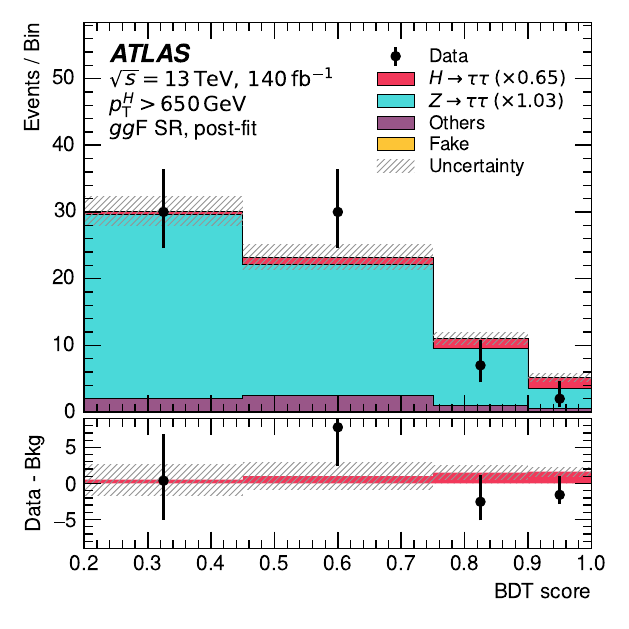} &
\includegraphics[scale=0.54]{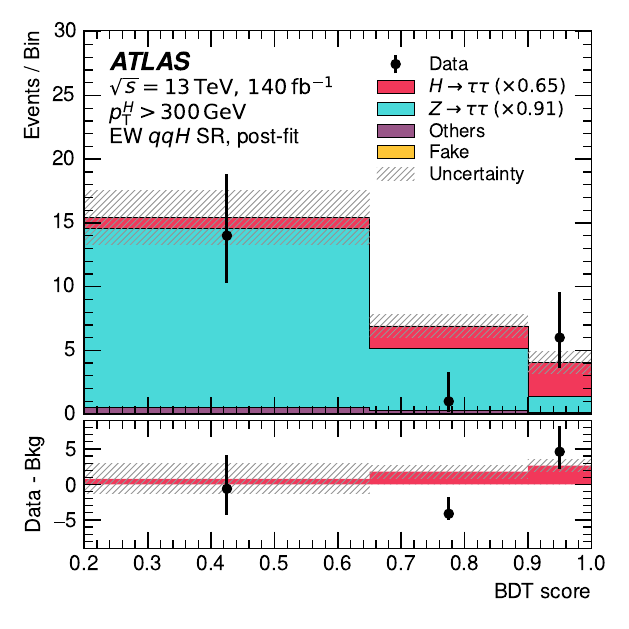} \\
(c) & (d) \\
\includegraphics[scale=0.54]{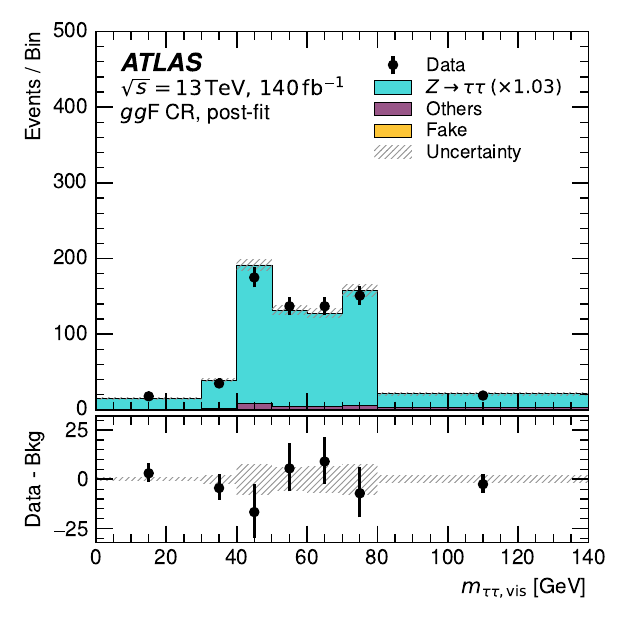} &
\includegraphics[scale=0.54]{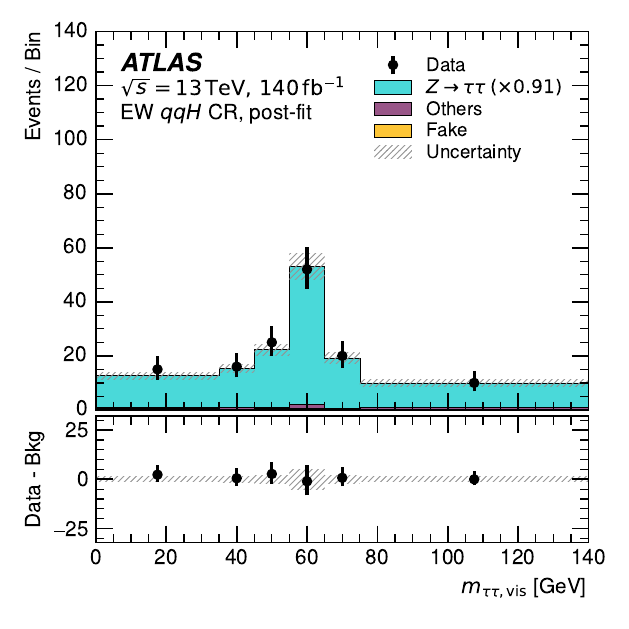} \\
(e) & (f) \\
\end{tabular}
\caption{Signal and control region distributions for the Run~2 data after the likelihood fit to measure the cross-section of \Htautau with $\pTH > 300$~\GeV: (a) $gg\text{F}(300\ \GeV < \pTH<450\ \GeV)$, (b) $gg\text{F}(450\ \GeV < \pTH< 650\ \GeV)$, (c) $gg\text{F}(\pTH>650\ \GeV)$, (d) EW qqH region, (e) $gg$F-like $Z\rightarrow\tau\tau$ control region and (f) EW qqH-like $Z\rightarrow\tau\tau$ control region. The signal is normalised by the signal strength value. The $\zttplusjets$ contribution is normalised by two factors, one for VBF-like events and another one for $gg$F-like events. `Others' refer to all minor backgrounds estimated with simulation. The lower panels show the differences between data and background with the solid bands representing the signal contribution. The uncertainty bars in the top (bottom) panel represent the statistical uncertainty on the data (difference between data and background).  The uncertainty bands include both statistical and systematic uncertainties from the fit used to extract the results. %
}
\label{fig:postfitrun2}
\end{center}
\end{figure}

\begin{figure}[h!]
\begin{center}
\begin{tabular}{cc}
\includegraphics[scale=0.54]{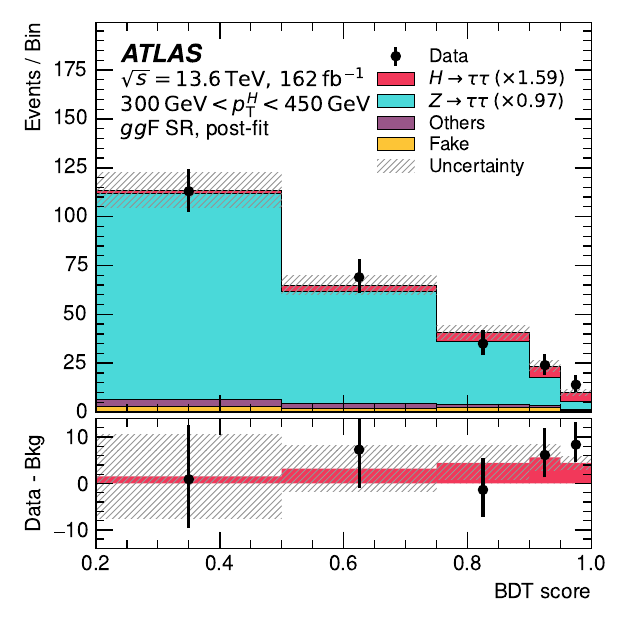} &
\includegraphics[scale=0.54]{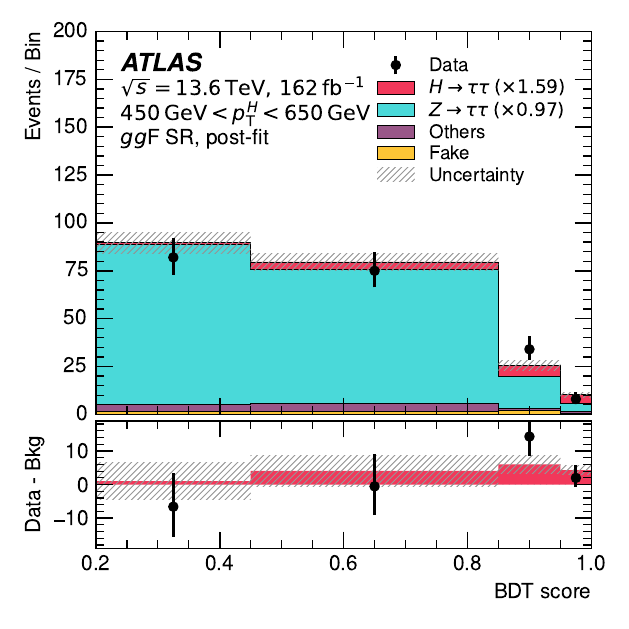} \\
(a) & (b) \\
\includegraphics[scale=0.54]{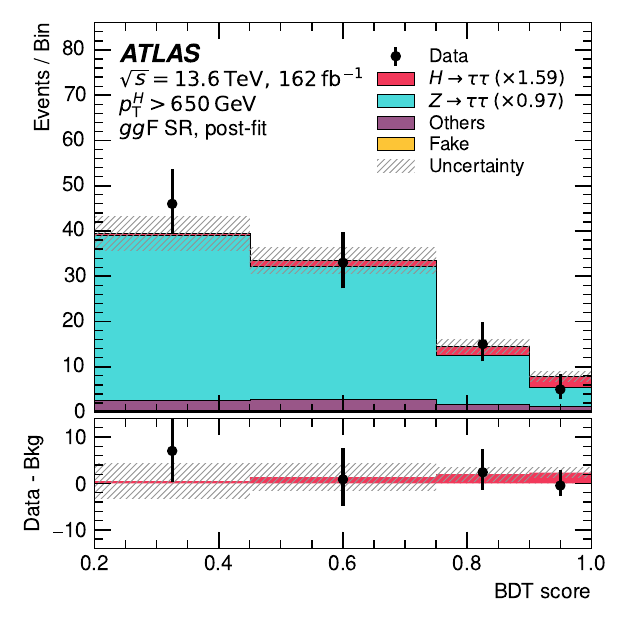} &
\includegraphics[scale=0.54]{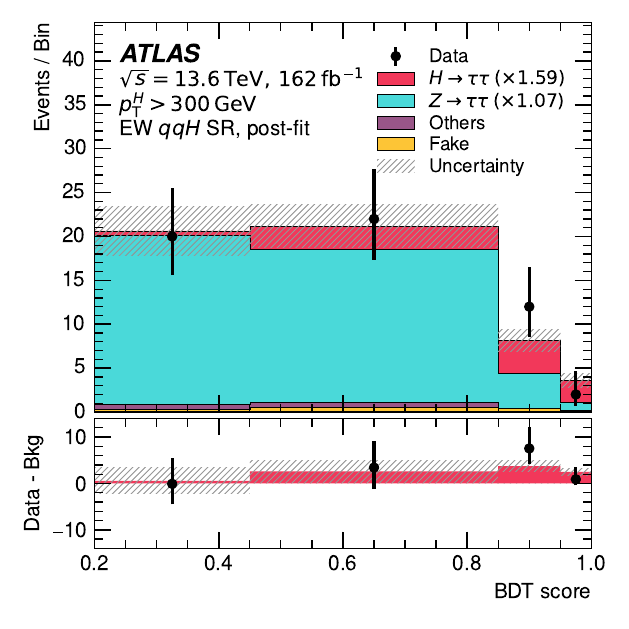} \\
(c) & (d) \\
\includegraphics[scale=0.54]{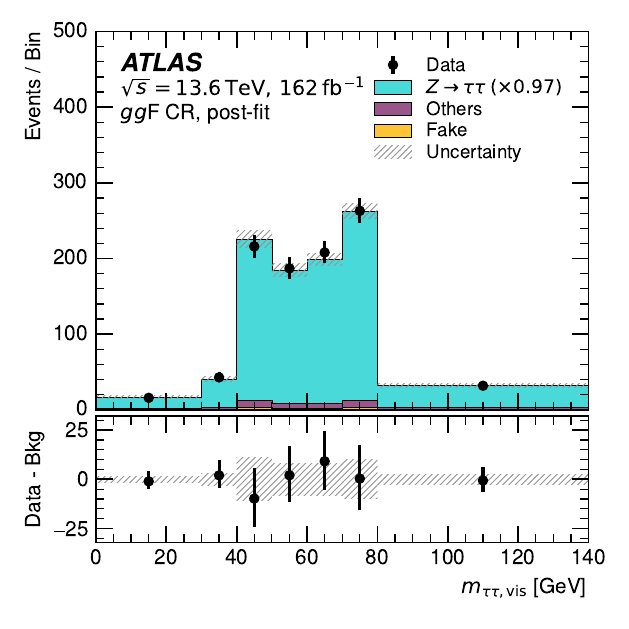} &
\includegraphics[scale=0.54]{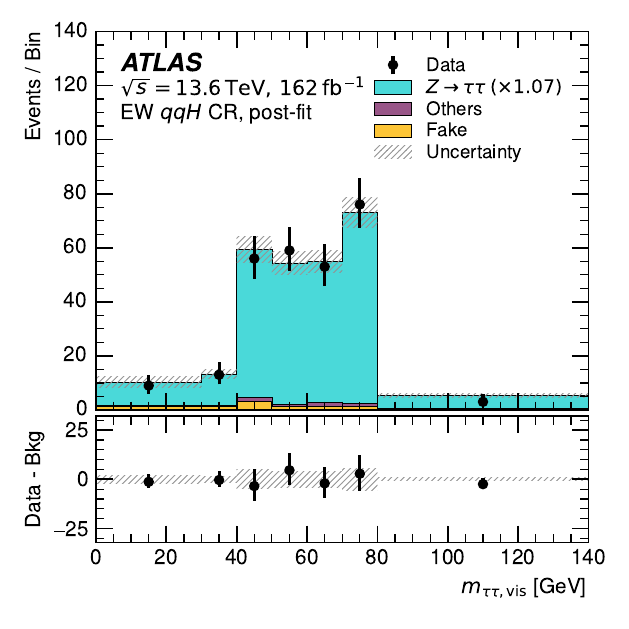} \\
(e) & (f) \\
\end{tabular}
\caption{Signal and control region distributions for the Run~3 data after the likelihood fit to measure the cross-section of \Htautau with $\pTH > 300$~\GeV: (a) $gg\text{F}(300\ \GeV < \pTH<450\ \GeV)$, (b) $gg\text{F}(450\ \GeV < \pTH< 650\ \GeV)$, (c) $gg\text{F}(\pTH>650\ \GeV)$, (d) EW qqH region, (e) $gg$F-like $Z\rightarrow\tau\tau$ control region and (f) EW qqH-like $Z\rightarrow\tau\tau$ control region. The signal is normalised by the signal strength value. The $\zttplusjets$ contribution is normalised by two factors, one for VBF-like events and another one for $gg$F-like events. `Others' refer to all minor backgrounds estimated with simulation. The lower panels show the differences between data and background with the solid bands representing the signal contribution. The uncertainty bars in the top (bottom) panel represent the statistical uncertainty on the data (difference between data and background).  The uncertainty bands include both statistical and systematic uncertainties from the fit used to extract the results.  %
}
\label{fig:postfitrun3}
\end{center}
\end{figure}

\begin{figure}
\begin{center}
\includegraphics[width=0.60\linewidth]{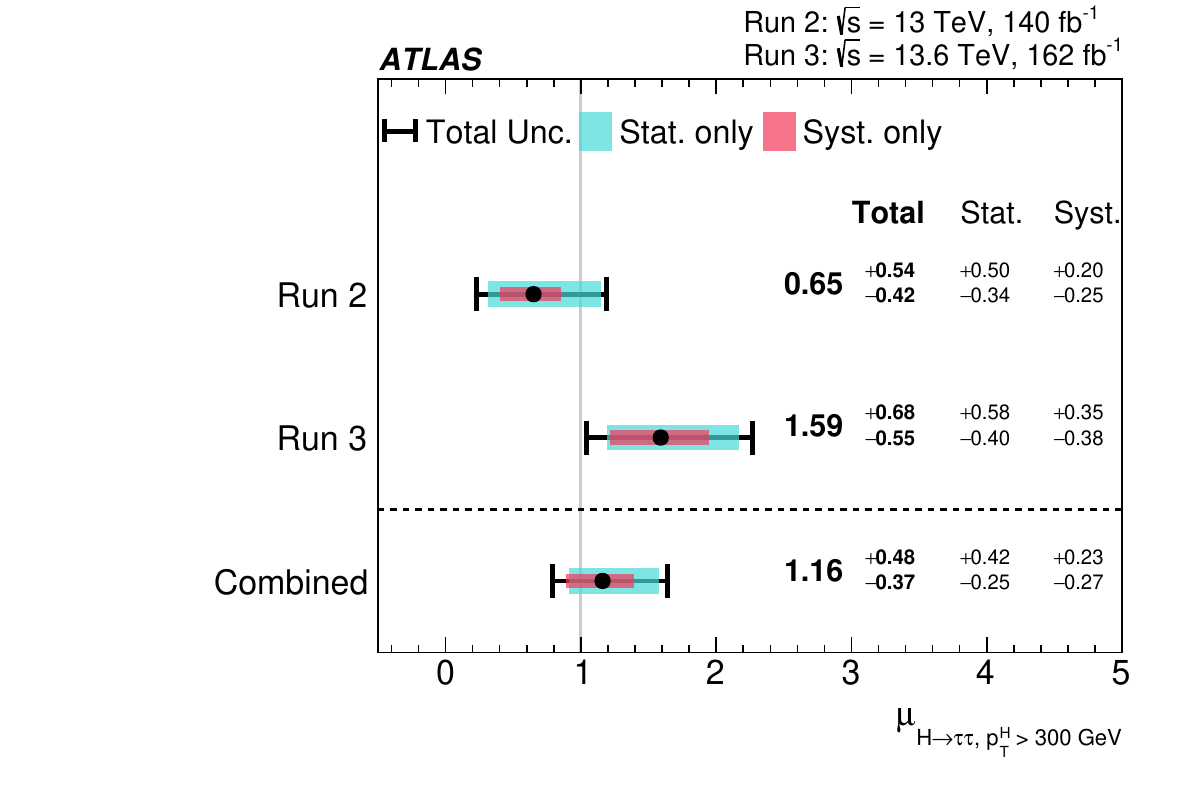}
\end{center}
\caption{Signal strength values determined in the likelihood fit to measure the cross-section of \Htautau with $\pTH > 300$~\GeV for the Run~2, Run~3 and combined data samples.}
\label{fig:1poires}
\end{figure}

\begin{table}[h!]
\caption{ Observed and expected signal significance determined from the likelihood fit to measure the cross-section of \Htautau with $\pTH > 300$~\GeV for the Run~2, Run~3 and combined data samples.}
\begin{center}
\begin{tabular}{%
@{}l%
S%
S%
@{}%
}
\hline
\hline
\multirow{2}{*}{}      & \multicolumn{2}{c}{ Signal significance for $\pTH > 300$~\GeV[$\sigma$] } \\
& \multicolumn{1}{c}{Observed} & \multicolumn{1}{c}{Expected}             \\
\hline
Run~2      &  1.7 & 2.1 \\\hline
Run~3      &  3.6 & 2.6 \\\hline
Combined   &  3.8 & 3.3 \\
\hline
\hline
\end{tabular}
\end{center}
\label{tab:sgn_values}
\end{table}

\cref{fig:stxspois} shows the STXS measured cross-sections, with respect to the corresponding SM predictions.
The compatibility of the fit with the SM expectation is 52\% (70\%) for the Run~2 (Run~3) data.
All the results are compatible with unity within one $\sigma$ except for Run~2, with a signal deficit for $gg$F-like events with $\pTH > 650$~\GeV.
For this $gg$F \pt bin, combining the Run~2 and Run~3 data samples, the observed (expected) upper limit on the cross-section is 4.5 (6.5) times the SM expectation  at 95\% confidence level.
Differences relative to the SM prediction between Run~2 and Run~3 data can be attributed to statistical fluctuations.

\begin{figure}[h!]
\begin{center}
\includegraphics[width=0.70\linewidth]{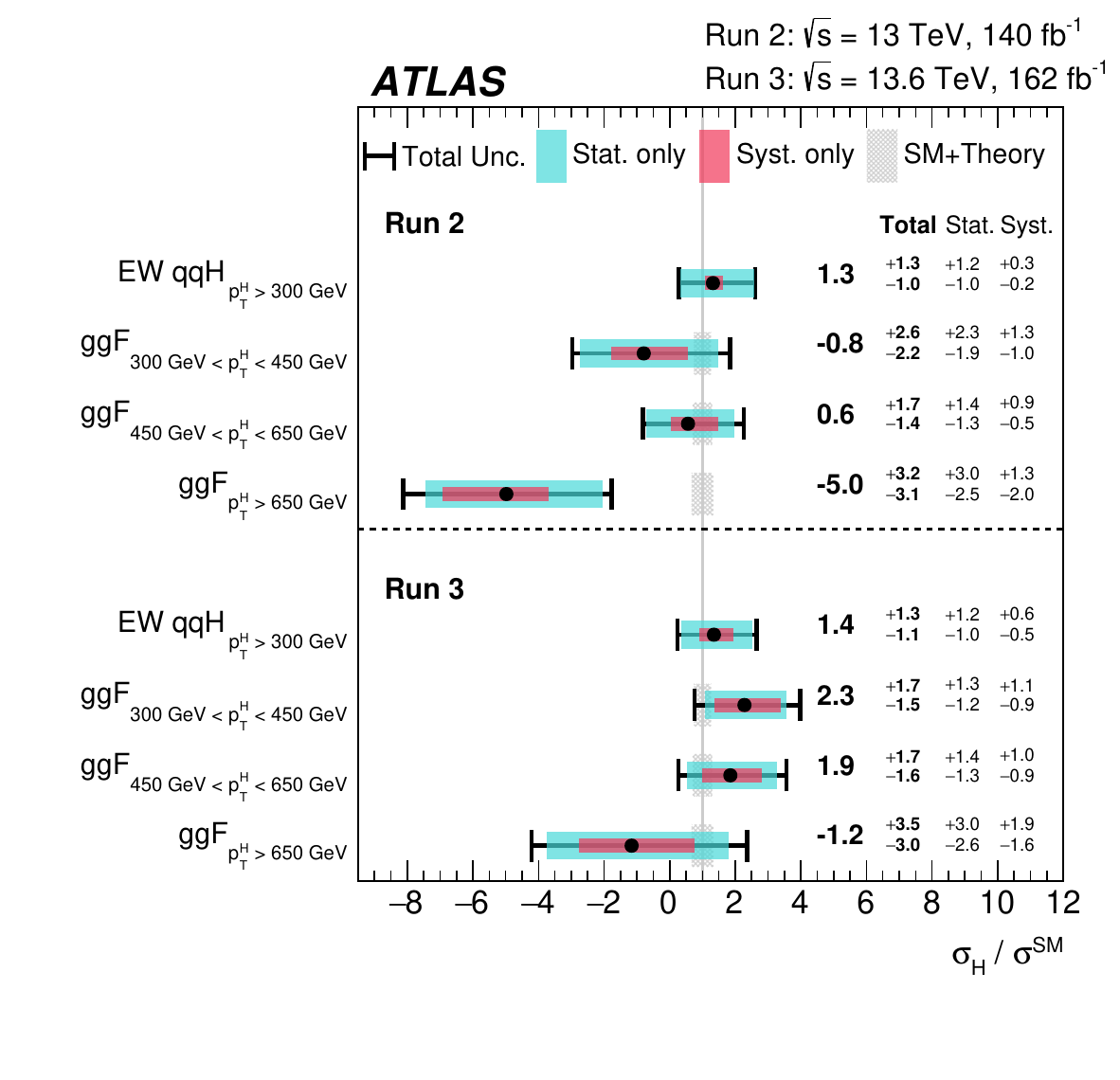}
\end{center}
\caption{STXS cross-section measurements normalised by the expected SM cross-section for the $gg\text{F}(300\ \GeV < \pTH<450\ \GeV)$, $gg\text{F}(450\ \GeV < \pTH< 650\ \GeV)$, $gg\text{F}(\pTH>650\ \GeV)$, and EW $qqH$ bins for the Run~2 and Run~3 data.}
\label{fig:stxspois}
\end{figure}

Finally, \cref{fig:differential} shows the measured differential cross-sections, with respect to the corresponding SM predictions.
The compatibility of the fit with the SM expectation is 33\% (56\%) for the Run~2 (Run~3) data.
Compared with the $H\rightarrow bb$ measurement in Ref.~\cite{HIGP-2024-01} for the $450\ \GeV < \pTH< 650\ \GeV$ interval, the two analyses find compatible upward cross-section values relative to the SM expectations in the Run~3 data. For the Run~2 data, an upward (downward) cross-section value relative to the SM expectation is found in $H\rightarrow bb$ ($H\rightarrow \tau\tau$).

\begin{figure}[h!]
\begin{center}
\includegraphics[width=0.60\linewidth]{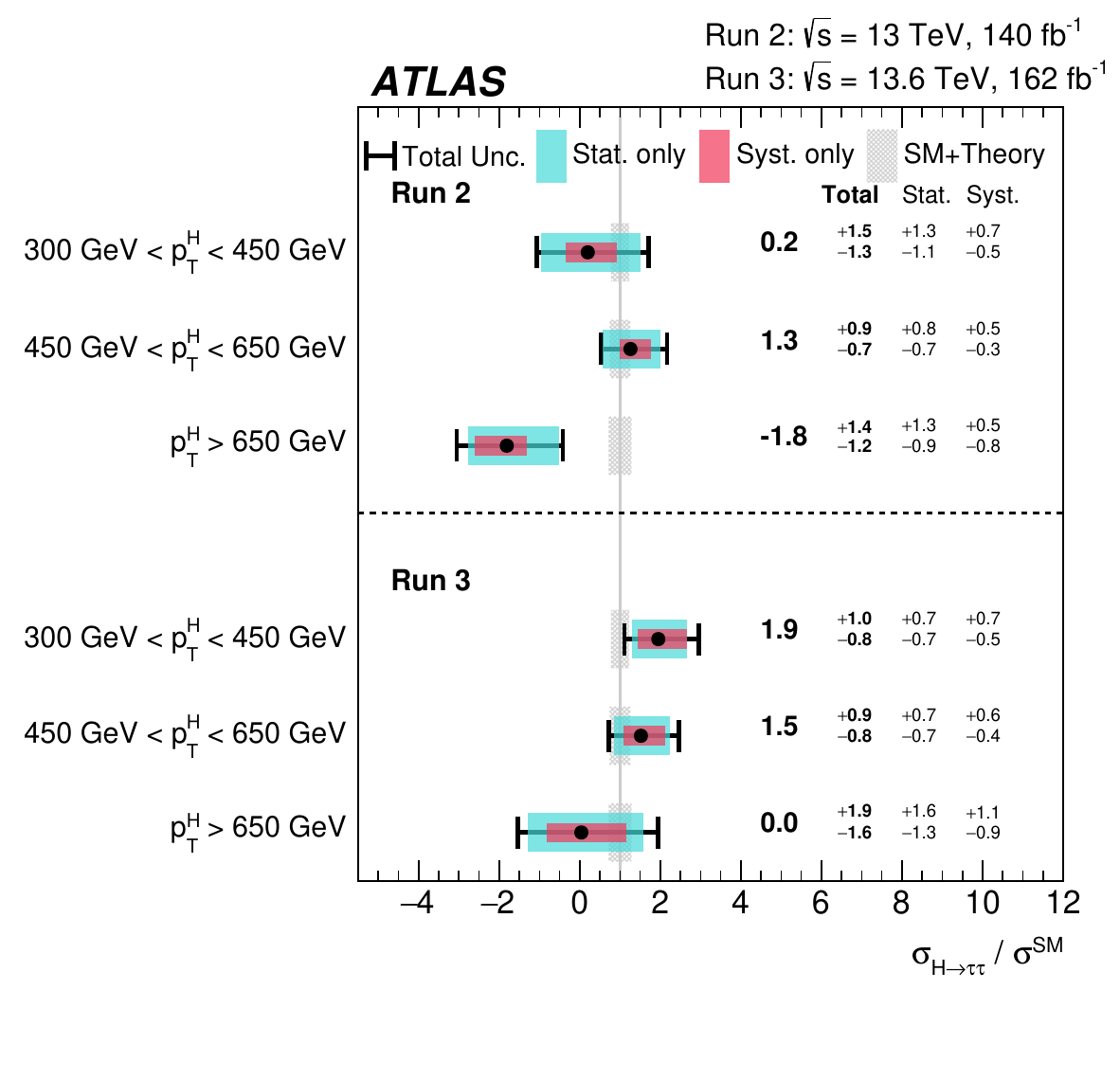}
\end{center}
\caption{Cross-section relative to SM expectation values  determined in the likelihood fit for the differential measurement for $300\ \GeV < \pTH<450\ \GeV$, $450\ \GeV < \pTH< 650\ \GeV$, and $\pTH>650\ \GeV$ bins for the Run~2 and Run~3 data.}
\label{fig:differential}
\end{figure}


%

%
%
\FloatBarrier

\section{Conclusion}
\label{sec:conclusion}

%
Cross-section measurements of boosted Higgs boson production in final states with pairs of hadronically decaying $\tau$-leptons are presented.
The phase space of interest corresponds to a Higgs boson transverse momentum of $\pTH > 300$~\GeV.
To conduct the measurements, dedicated techniques for boosted di-$\tauhad$ reconstruction and identification are developed.
The analysis uses a data sample corresponding to the Run 2 (Run 3) data that was collected in 2015--2018 (2022--2024), at a centre-of-mass energy of $\sqrt{s}$ = 13 (13.6)~\TeV, and corresponding to an integrated luminosity of 140 (162) $\ifb$.
Combining both data samples and considering only signal events with $\pTH > 300$~\GeV, the  observed (expected) signal significance is 3.8 (3.3) $\sigma$ over the background-only hypothesis, hence providing evidence for Higgs boson production in the boosted regime in the di-$\tau$ final state.
Cross-section measurements are also reported within the simplified template cross-section framework and differentially relative to $\pTH$.
All results are compatible with the Standard Model predictions.


%
%
%
%
%
\section*{Acknowledgements}
%

%

%
%

%
%

We thank CERN for the very successful operation of the LHC and its injectors, as well as the support staff at
CERN and at our institutions worldwide without whom ATLAS could not be operated efficiently.

The crucial computing support from all WLCG partners is acknowledged gratefully, in particular from CERN, the ATLAS Tier-1 facilities at TRIUMF/SFU (Canada), NDGF (Denmark, Norway, Sweden), CC-IN2P3 (France), KIT/GridKA (Germany), INFN-CNAF (Italy), NL-T1 (Netherlands), PIC (Spain), RAL (UK) and BNL (USA), the Tier-2 facilities worldwide and large non-WLCG resource providers. Major contributors of computing resources are listed in Ref.~\cite{ATL-SOFT-PUB-2026-001}.

We gratefully acknowledge the support of ANPCyT, Argentina; YerPhI, Armenia; ARC, Australia; BMWFW and FWF, Austria; ANAS, Azerbaijan; CNPq and FAPESP, Brazil; NSERC, NRC and CFI, Canada; CERN; ANID, Chile; CAS, MOST, NSFC and FRFCU, China; Minciencias, Colombia; MEYS CR, Czech Republic; DNRF and DNSRC, Denmark; IN2P3-CNRS and CEA-DRF/IRFU, France; SRNSFG, Georgia; BMFTR, HGF and MPG, Germany; GSRI, Greece; RGC and Hong Kong SAR, China; ICHEP and Academy of Sciences and Humanities, Israel; INFN, Italy; MEXT and JSPS, Japan; CNRST, Morocco; NWO, Netherlands; RCN, Norway; MNiSW, Poland; FCT, Portugal; MNE/IFA, Romania; MSTDI, Serbia; MSSR, Slovakia; ARIS and MVZI, Slovenia; DSI/NRF, South Africa; MICIU/AEI, Spain; SRC and Wallenberg Foundation, Sweden; SERI, SNSF and Cantons of Bern and Geneva, Switzerland; NSTC, Taipei; TENMAK, T\"urkiye; STFC/UKRI, United Kingdom; DOE and NSF, United States of America.

Individual groups and members have received support from BCKDF, CANARIE, CRC and DRAC, Canada; CERN-CZ, FORTE and PRIMUS, Czech Republic; COST, ERC, ERDF, Horizon 2020 and Marie Sk{\l}odowska-Curie Actions, European Union; Investissements d'Avenir Labex, Investissements d'Avenir Idex and ANR, France; DFG and AvH Foundation, Germany; Herakleitos, Thales and Aristeia programmes co-financed by EU-ESF and the Greek NSRF, Greece; BSF-NSF and MINERVA, Israel; NCN and NAWA, Poland; La Caixa Banking Foundation, CERCA and AGAUR programs from Generalitat de Catalunya and PROMETEO and GenT Programmes Generalitat Valenciana, Spain; G\"{o}ran Gustafssons Stiftelse, Sweden; The Royal Society and Leverhulme Trust, United Kingdom; Eric and Wendy Schmidt Fund for Strategic Innovation, United States of America.

In addition, individual members wish to acknowledge support from Chile: Agencia Nacional de Investigaci\'on y Desarrollo (ANID FONDECYT reg. 1230987, FONDECYT 1230812, FONDECYT 1240864, Fondecyt 3240661, Fondecyt Regular 1240721); China: Fundamental Research Funds for the Central Universities (010-63263105), Chinese Ministry of Science and Technology (MOST-2023YFA1605700, MOST-2023YFA1609300), National Natural Science Foundation of China (NSFC 12275265, NSFC-W2543005); Czech Republic: Czech Science Foundation (GACR - 24-11373S), Ministry of Education Youth and Sports (ERC-CZ-LL2327, FORTE CZ.02.01.01/00/22\_008/0004632); EU: H2020 European Research Council (ERC - 101002463); European Union: European Research Council (BARD No. 101116429, ERC - 101219398, ERC 101089007), European Regional Development Fund (HE COFUND GA No.101081355, ERDF), Marie Sklodowska-Curie Actions (GAP-101168829); France: Agence Nationale de la Recherche (ANR-21-CE31-0013, ANR-22-EDIR-0002, ANR-24-CE31-0504-01); Germany: Deutsche Forschungsgemeinschaft (DFG - 469666862); China: Research Grants Council (GRF); Italy: Istituto Nazionale di Fisica Nucleare (LHC-MIUR - 28003/2025), Ministero dell'Università e della Ricerca (NextGenEU I53D23000820006 M4C2.1.1, SOE2024\_0000023); Japan: Japan Society for the Promotion of Science (JSPS KAKENHI  JP25H0063, JSPS KAKENHI JP22H04944, JSPS KAKENHI JP24K23939, JSPS KAKENHI JP24KK0251, JSPS KAKENHI JP25H00650, JSPS KAKENHI JP25H01291, JSPS KAKENHI JP25K01011, JSPS KAKENHI JP25K01023, JSPS KAKENHI JP25KK0047); Poland: Polish National Science Centre (NCN 2021/42/E/ST2/00350, NCN OPUS 2023/51/B/ST2/02507, NCN OPUS nr 2022/47/B/ST2/03059, NCN UMO-2019/34/E/ST2/00393, UMO-2022/47/O/ST2/00148, UMO-2023/49/B/ST2/04085, UMO-2023/51/B/ST2/00920, UMO-2024/53/N/ST2/00869); Spain: Agència de Gestió d'Ajuts Universitaris i de Recerca. (AGAUR - 2023 BP 00141), Ministry of Science and Innovation (RYC2019-028510-I, RYC2020-030254-I, RYC2021-031273-I, RYC2022-038164-I), Ministerio de Ciencia, Innovación y Universidades/Agencia Estatal de Investigaci\'on (EU NextGenerationEU (PRTR-C17.I1), PID2022-142604OB-C22); Sweden: Carl Trygger Foundation (Carl Trygger Foundation CTS 22:2312), Swedish Research Council (Swedish Research Council 2023-04654, VR 2021-03651, VR 2022-03845, VR 2022-04683, VR 2023-03403, VR 2024-05451, VR 2025-05940), Knut and Alice Wallenberg Foundation (KAW 2023.0366); United Kingdom: The Binks Trust, Royal Society (NIF-R1-231091); United States of America: U.S. Department of Energy (ECA DE-AC02-76SF00515), John Templeton Foundation (John Templeton Foundation 63206), Neubauer Family Foundation.

%
%


%
%
%
%
%
%
%

%
%

%
%
\clearpage
\printbibliography
\clearpage
 
\begin{flushleft}
\hypersetup{urlcolor=black}
{\Large The ATLAS Collaboration}

\bigskip

\AtlasOrcid[0000-0002-6665-4934]{G.~Aad}$^\textrm{\scriptsize 102}$,
\AtlasOrcid[0000-0001-7616-1554]{E.~Aakvaag}$^\textrm{\scriptsize 17}$,
\AtlasOrcid[0000-0002-5888-2734]{B.~Abbott}$^\textrm{\scriptsize 121}$,
\AtlasOrcid[0000-0002-0287-5869]{S.~Abdelhameed}$^\textrm{\scriptsize 83b}$,
\AtlasOrcid[0000-0002-1002-1652]{K.~Abeling}$^\textrm{\scriptsize 54}$,
\AtlasOrcid[0000-0001-5763-2760]{N.J.~Abicht}$^\textrm{\scriptsize 48}$,
\AtlasOrcid[0000-0002-8496-9294]{S.H.~Abidi}$^\textrm{\scriptsize 30}$,
\AtlasOrcid[0009-0003-6578-220X]{M.~Aboelela}$^\textrm{\scriptsize 44}$,
\AtlasOrcid[0000-0002-9987-2292]{A.~Aboulhorma}$^\textrm{\scriptsize 36e}$,
\AtlasOrcid[0000-0001-5329-6640]{H.~Abramowicz}$^\textrm{\scriptsize 154}$
% ...
\bigskip
\\

% ...
$^{17}$Department for Physics and Technology, University of Bergen, Bergen; Norway.\\
% ...
$^{30}$Physics Department, Brookhaven National Laboratory, Upton NY; United States of America.\\
% ...
$^{36}$$^{(a)}$Facult\'e des Sciences Ain Chock, Universit\'e Hassan II de Casablanca;$^{(b)}$Facult\'{e} des Sciences, Universit\'{e} Ibn-Tofail, K\'{e}nitra;$^{(c)}$Facult\'e des Sciences Semlalia, Universit\'e Cadi Ayyad, LPHEA-Marrakech;$^{(d)}$LPMR, Facult\'e des Sciences, Universit\'e Mohamed Premier, Oujda;$^{(e)}$Facult\'e des sciences, Universit\'e Mohammed V, Rabat;$^{(f)}$Institute of Applied Physics, Mohammed VI Polytechnic University, Ben Guerir; Morocco.\\
% ...
$^{44}$Physics Department, Southern Methodist University, Dallas TX; United States of America.\\
% ...
$^{48}$Fakult\"{a}t Physik, Technische Universit{\"a}t Dortmund, Dortmund; Germany.\\
% ...
$^{54}$II. Physikalisches Institut, Georg-August-Universit\"{a}t G\"ottingen, G\"ottingen; Germany.\\
% ...
$^{83}$$^{(a)}$Khalifa University of Science and Technology, Abu Dhabi;$^{(b)}$New York University Abu Dhabi, Abu Dhabi;$^{(c)}$United Arab Emirates University, Al Ain;$^{(d)}$University of Sharjah, Sharjah; United Arab Emirates.\\
% ...
$^{102}$CPPM, Aix-Marseille Universit\'e, CNRS/IN2P3, Marseille; France.\\
% ...
$^{121}$Homer L. Dodge Department of Physics and Astronomy, University of Oklahoma, Norman OK; United States of America.\\
% ...
$^{154}$Raymond and Beverly Sackler School of Physics and Astronomy, Tel Aviv University, Tel Aviv; Israel.\\
% ...
$^{a}$ Also at Affiliated with an institute formerly covered by a cooperation agreement with CERN.\\
$^{b}$ Also at An-Najah National University, Nablus; Palestine.\\
$^{c}$ Also at Borough of Manhattan Community College, City University of New York, New York NY; United States of America.\\
$^{d}$ Also at Center for Interdisciplinary Research and Innovation (CIRI-AUTH), Thessaloniki; Greece.\\
$^{e}$ Also at Centre of Physics of the Universities of Minho and Porto (CF-UM-UP); Portugal.\\
$^{f}$ Also at CERN, Geneva; Switzerland.\\
% ...

\end{flushleft}


%

%
%

%

%
%
%
%

\end{document}